\documentclass[11pt,twoside,a4paper]{article}
\usepackage[affil-it]{authblk}
\usepackage{amssymb}
\usepackage{amsmath}
\usepackage{tensor}
\usepackage{color}
\usepackage{tikz}
\usetikzlibrary{matrix}
\usepackage[margin=2.7 cm]{geometry}
\usepackage{polski}
\usepackage[cp1250]{inputenc}
\usepackage[polish,english]{babel}
\usepackage[toc,page]{appendix}
\usepackage{amsthm}
\usepackage{stmaryrd}

\usepackage{hyperref}

\input xy
\xyoption{all} \tolerance=500

\def\sT{\mathsf T}
\newdir{ (}{{}*!/-5pt/@^{(}}
\newdir{|>}{!/4,5pt/@{|}*:(1,-.2)@^{>}*:(1,+.2)@_{>}}

\def\mg{\mathfrak g}

\def\d{\mathrm d}

\def\sV{\mathsf V}

\def\sJ{\mathsf J^1}
\def\sJd{\mathsf J^\dagger}

\def\J2{\mathsf J^2}
\def\sj{\mathsf j^1}
\def\ind{\indices}
\def\blue{\color{blue}}
\def\red{\color{red}}

\def\Ad{\mathrm{Ad}}
\def\ad{\mathrm{ad}}

\def\ddt{\frac{\mathsf d}{\mathsf dt}_{|t=0}}

\def\cV{\mathcal V}
\def\cF{\mathcal F}
\def\C2{\mathbb C^2}
\def\barP{\overline{\mathcal P}}
\def\Aff{\mathsf{Aff}}
\def\wdt{\widetilde }

\newtheorem{theorem}{Theorem}
\newtheorem{lemma}{Lemma}

\title{The Tulczyjew triple for gauge field theories}

\begin{document}

\author{Marcin Zaj\k{a}c}
\date{}
\maketitle

\vskip -1.0cm

\centerline{Department of Mathematical Methods in Physics,}
\centerline{Faculty of Physics. University of Warsaw,}
\centerline{ul. Pasteura 5, 02-093 Warsaw, Poland.}
\centerline{marcin.zajac@fuw.edu.pl}

\vskip 1.0cm

\begin{abstract}
In this article we construct and discuss a new rigorous geometric formalism for gauge field theories. The basis of our work is the notion of the Tulczyjew triple, a geometric structure which successfully solved numerous problems in mathematical description of mechanics and classical field theory. In particular, we  construct a Tulczyjew triple for gauge field theories and reduce it for systems that depend only on the value of a connection and curvature instead of the entire first jet of the gauge field. We also introduce new geometric structures such as the vector-affine product of bundles and analyse the connection bundle from a new perspective.
Finally, we apply the derived formalism to Yang-Mills theory.

\end{abstract}

\tableofcontents

\section{Introduction}

{\bf  The Tulczyjew triples.} Geometrical tools of the classical field theory were developed as a generalisation of ideas coming from classical mechanics. In most textbooks and papers, analytical mechanics is based on variational calculus and its main objective is Euler- Lagrange equation. Analogously, using variational principle, one may derive field equations in classical field theories.
However, if one wants to describe field theory geometrically, it turns out that there is a surprisingly large variety of approaches. The most common one is to extend the symplectic formalism from mechanics to field theory, which leads to polysymplectic structures \cite{CG}, $k$-symplectic structures \cite{AA,MP,LV}, $k$-almost cotangent structures \cite{LMS1,LMS2} and multisymplectic structures \cite{JK,HS,F2,F3}. However, each of these approaches has significant limits, which immediately appear when one wants to consider nonregular systems, reduction with respect to symmetries or inclusion of contraints \cite{Bergman}. %Since we can not present all of them in detail, we will just mention that each of them has its defects and virtues. 

In 70's W. M. Tulczyjew suggested a new understanding of variational calculus in mechanics and field theories. In his numerous works (e.g \cite{WMT1,WMT2,WMT3,WMT4}) he presented a formalism in which, in contrast to most textbooks, the crucial object is the phase dynamics instead of Euler-Lagrange equation. This formalism was later called a {\it Tulczyjew triple}. It has been recently recognised by many theoretical physicist and mathematicians. It provides more general and complete description of a mechanical system and is much simpler on the conceptual level than the traditional one.

Tulczyjew triple is a very useful commutative diagram built on maps that are essential in Lagrangian and Hamiltonian description of physical systems. In fact the name {\it Tulczyjew triple} refers not to one diagram but a collection of diagrams adapted to various physical situations. The very first triple introduced by Tulczyjew in his numerous works (e.g \cite{WMT3,WMT4,WMT6}) served for autonomous analytical mechanics. It was then adapted and generalised for time-dependent mechanics, mechanics on algebroids \cite{GG,GGU,ZG}, %constrained mechanics on Dirac algebroids \cite{GG2}, 
for field theory \cite{KG}, for higher order systems \cite{GV} etc. The concept of Tulczyjew triple also points to the certain philosophy of interpreting concepts of variational calculus within physical theories.

The main advantage of the approach developed by Tulczyjew and his collaborators is its generality. For example, using the Tulczyjew triple for autonomous mechanics we can derive the phase equations for systems with singular Lagrangians and understand properly the Hamiltonian description of such systems. One can even discuss systems with more general generating objects than just a Lagrangian function, e.g. systems described by family of Lagrangians or a Lagrangian function defined on a submanifold. Another advantage of the Tulczyjew’s approach is its flexibility. Being based on well-defined general principles, it can be easily adapted to different settings \cite{KG,GG,GV,LMM}. Finally, we do not postulate {\it ad hoc} the ingredients of the theory, but obtain them as unavoidable consequences of the variational calculus \cite{KG}.

{\bf Gauge field theories.} Within our work we will be interested in gauge field theories. From the mathematical point of view gauge field is represented by a connection in a principal bundle $\pi:P\to M$ with the structure group $G$. Therefore, gauge theories may be, to some extent, reduced to the geometry of principal bundles. {\it A gauge transformation} of the principal bundle $P$ is an equivariant diffeomorphism $\Phi:P\to P$ preserving the projection on the base manifold $M$. By equivariance we mean that the condition $\Phi(pg)=\Phi(p)g$ is satisfied for each $p\in P$, $g\in G$. The set of gauge transformations form an infinite-dimensional group with a composition of maps as a group multiplication. We denote this group by $\mathcal G(P)$. We say that  $\mathcal G(P)$ is a symmetry group of a given theory if its action functional is invariant under the transformations given by $\mathcal G(P)$ \cite{JMFF,SG2}.

The first historically discovered gauge theory was classical electrodynamics formulated by J. C. Maxwell \cite{JCM}. The existence of a gauge symmetry within Maxwell equations initially did not appear to be a fact of huge importance. Only the papers of Weyl on the unification of electrodynamics with general relativity consciously introduced the notion of a local symmetry to theoretical physics \cite{HWeyl}. In his seminal paper \cite{HWeyl} Weyl introduced a term ,,gauge transformation", and in particular gauge invariance (ger. "eichinvarianz"). However, perhaps the greatest success of gauge theories came in 1954, when C. H. Yang and R. Mills introduced non-abelian gauge theories to describe strong interaction confining nucleons in atomic nucleus \cite{YangMills}. These theories came to be known in the literature as {\it Yang-Mills theories}, after their inventors. Ever since then Yang-Mills theories and gauge theories in general became one of the main objects of study in theoretical physics \cite{CLF,DV}. Arguably, the most important discovery within this field was the emergence of the Standard Model of particle physics.  %Prawdopodobnie najbardziej spektakularnym sukcesem tej klasy teorii było powstanie Modelu Standardowego cząstek elementarnych.

{\bf Main goals and results.} The first goal of our paper is to find a mathematical formalism, which will allow us to describe in a simple way the dynamics of gauge fields both on the Lagrangian and Hamiltonian side. In this part of our work we mainly rely on the formalism of the Tulczyjew triple, which is described in detail in a section 3. One of the main advantages of the Tulczyjew description is that it does not require any regularity of the Lagrangian to find the dynamics of the system and to pass to the Hamiltonian side. This feature is particularly important in the context of field theory. 
While in mechanics regular sytems constitute a vast majority of the physically interesting models, in field theory (especially gauge theories) most of the systems is nonregular. Whatsmore, many of these teories are naturally theories with constraints. Since the Tulczyjew formalism turned out to be very efficient in application to both constrained systems and nonregular systems, it seems to be a perfect tool to analyse such problems.

%{\blue -------------------------------------- paragraph above is ok ------------------------------------}

In the next step we want to reduce the derived Tulczyjew triple with respect to internal symmetries appearing in gauge theories. Let us stress here that in the current paper we do not consider the problem of gauge symmetries reduction. We plan to work on this issue in a separate paper that is aimed as the continuation of this one. In particular, we will be interested now in a following problem. The general formalism of the classical field theory is based on the geometry of jet bundles. The space of fields is represented by a fiber bundle $E\to M$, where $E$ is the space of values of the field. To find the dynamics of a system one usually takes the Lagrangian, which is a map
$$ L:\sJ E\to\Omega^m,   $$
where $\Omega^m$ is the space of $m$-covectors on the $m$-dimensional manifold $M$.
However, in gauge theories a Lagrangian usually does not depend on the entire first jet of a connection but only on the value of the field and on its curvature. This is the case for instance of such important theories as classical electrodynamics or Yang-Mills and Yang-Mills-Higgs theories. Therefore, a natural question arises, how to describe geometrically the projection associated with the passage from the first jet of the connection to the curvature of this connection in a given point. This kind of projection should imply a reduction of the entire structure of gauge theories depending only on the value of the field and its curvature. In this paper we discuss this reduction in detail and, as a result, we obtain a reduced Tulczyjew triple for gauge fields.  

An additional result of the paper is the further development of geometric structures associated with gauge theories. In our formulation the space of gauge fields is a bundle $C\to M$, where $C=\sJ P/G$. In section 5 we provide a meticulous analysis of the connection bundle in the realm of jet bundles and group actions. 
In order to do that we introduce a new kind of product of bundles in section 4, which we call a {\it vector-affine product}.
The main result of this part of our work is Theorem 1 in subsection 5.3, which shows that $\sJ C$ has a natural structure of the vector-affine product. 
Let us mention that although the jet bundle geometry is a natural area to develop classical field theory, the literature concerning gauge theories in the jet bundle setting is extremely modest \cite{SG2,SG3}. In particular, in \cite{SG2} Sardanashvily mentions a result similar to Theorem 1 but provides a wrong formula for it. We will discuss this issue in detail in section 5.3. %Finally, we show {\blue yang mills }.

%{\red Here write something more. Maybe add something from Significance of research. That we analyse bundle JC etc... Write more about results, for instance vector affine product and Sardanaschvhily. Write that we want to start a rigorous treatment of gauge theories and in the futue we will deal with gauge transformations.}

{\bf Outline of the paper.} 
The paper is organized as follows. Section 2 contains a brief recalling of the basic tools used in the geometric description of classical field theory of first order and gauge theories. The main notions in this context are jet bundles and principal bundles. In section 3 we recall the concept of a Tulczyjew triple in mechanics and classical field theory of the first order. In section 4 we introduce a notion of the vector-affine bundles and we consider the tangent and cotangent bundle of such products. Section 5 contains the detailed discussion of the connection bundle in a bit different language than the one presented in most textbooks. Sections 6 and 7 are main results of the paper and they contain a reduction of the Tulczyjew formalism for gauge theories. %by means of tools derived in previous sections. 
The results of this reduction are summarised in section 8. In section 9 we apply the reduced Tulczyjew triple to Yang-Mills theory.

\section{Jet bundles and principal bundles}

\subsection{First order jet bundles}

Let us briefly recall the notion of first order jet spaces and their duals. We will follow the notation from \cite{KG}. For a more detailed discussion of the jet bundle geometry see e.g. \cite{Saund,EG1}.

Let $\pi :E\to M$ be a bundle with the total space of dimension $\mathrm{dim}  E=n+m$. We introduce in a domain $U\in M$ a local coordinate system $(x^i)^{n}_{i=0}$ on $M$. In field theory fields are represented by sections of a fibration $\pi$. The total space is the space of values of the field e.g vector field is a section of $\pi$ being a vector bundle, scalar field is a section of a trivial bundle $E=M\times\mathbb R$ or $E=M\times\mathbb C$, etc. On an open subset $V\subset E$ such that $\pi(V)=U$ we introduce local coordinates $(x^i,u\ind{^\alpha})$ adapted to the structure of the bundle. 

In $\sT E$ we have a vector subbundle $\sV E\to E$ consisting of those tangent vectors that are vertical with respect to the projection $\pi$, i.e. $\sT\pi(v_p)=0$ for $v_p\in\sV_pE$. We will also need its dual vector bundle $\sV^*E\to E$.

The space of first jets of sections of the bundle $\pi$ will be denoted by $\sJ E$.  By definition, the first jet
$\sj_m\phi$ of a section $\phi$ at the point $m\in M$ is an equivalence class of sections having the same value at the point $m$ and such that the spaces tangent to the graphs of the sections at the point $\phi(m)$ coincide. Therefore, there is a natural projection $\pi_{1,0}$ from the space $\sJ E$ onto the manifold $E$
$$\pi_{1,0}:\sJ E\to E: \quad  \mathsf j^1_m\phi\longmapsto\phi(m).    $$
Moreover, every jet $\sj_m\phi$ may be identified with a linear map $\sT\phi:\sT_mM\to\sT_{\phi(m)}E$. Linear maps coming from jets at the point $m$ form an affine subspace in a vector space $Lin(\sT_mM,\sT_eE)$ of all linear maps from $\sT_mM$ to $\sT_eE$. A map belongs to this subspace if composed with $\sT\pi$ gives identity. In a tensorial representation we have an inclusion
$$\sJ_eE\subset\sT^*_mM\otimes\sT_eE.$$ 
It is easy to check that the affine space $\sJ_eE$ is modelled on the vector space $\sT_m^*M\otimes\sV_eE$. Summarising, the bundle $\sJ E\to E$ is an affine bundle modelled on the vector bundle 
$$\pi^*(\sT^*M)\otimes_E\sV E\to E.$$
%The symbol $\pi^*(\sT^*M)$ denotes the pullback of the cotangent bundle $\sT^*M$ along to the projection $\pi$. 
In the following we will omit the symbol of the pullback in $\pi^*(\sT^*M)$ writing simply $\sT^*M\otimes_E\sT E$ and $\sT^*M\otimes_E\sV E$.

Using the adapted coordinates $(x^i,u\ind{^\alpha})$ in $V\subset E$, we can construct the induced coordinate system $(x^i,u\ind{^\alpha},u\ind{^\beta_j})$ on $\pi_{1,0}^{-1}(V)$ such that for any section $\phi$ given by $n$ functions $\phi^a(x^i)$ we have

$$u\ind{^\beta_j}(\phi^\alpha(x^i))=\frac{\partial\phi^\beta }{\partial x^j}(x^i(m)).$$
In the tensorial representation the first jet $\mathsf j^1_m\phi$ may be written as
$$\d x^i\otimes\frac{\partial}{\partial x^i}+\frac{\partial\phi^\alpha}{\partial x^j }(x^i(m))\d x^j\otimes\frac{\partial}{\partial u\ind{^\alpha}},    $$
where we have used local bases of sections of $\sT^*M$ and $\sT E$ coming from the chosen coordinates.    

%{\blue ------------------------------------- till now its ok ---------------------------------------}

We will introduce now the bundle which is dual to $\sJ E\to E$. Let us recall that each fiber $\sJ_eE$ is an affine space. We can consider a set of affine maps $\sJ_eE\to\mathbb R$, say $\mathsf{Aff}(\sJ_eE,\mathbb R)$, for each $e\in E$. Collecting $\mathsf{Aff}(\sJ_eE,\mathbb R)$ point by point we obtain the bundle of affine maps on $\sJ E$, namely $\mathsf{Aff}(\sJ E,\mathbb R)\to E$. From now on we will use the notation $\sJd E:=\mathsf{Aff}(\sJ E,\mathbb R)$. It is a vector bundle over $E$. If $(x^i,u\ind{^\alpha},u\ind{^\beta_j})$ are coordinates in $\sJ E$, then we introduce coordinates $(x^i,u\ind{^\alpha},r,\varphi\ind{^b_j})$ in $\sJd E$. The evaluation between $\sJ E$ and $\sJd E$ in coordinates reads
$$\sJd E\times_E\sJ E\to\mathbb R, \qquad \langle T_e,\mathsf j_e\psi\rangle=r+\varphi\ind{^b_j}y\ind{^j_b}.   $$

Let us recall that each affine map has the associated linear part. Since $\sJ_eE$ is modelled on the vector bundle $\sT^*_{\pi(e)}M\otimes_E\sV_eE$ the linear part of an affine map $T_e: \sJ_eE\to\mathbb R$ is an element of $\sT_{\pi(e)}M\otimes_E\sV^*_eE$. The bundle
$$ \mu:\sJd E\to\sT M\otimes_E\sV^*E$$
is an affine bundle, which projects an affine map onto its linear part. The model bundle of $\mu$ is a trivial bundle $\sT M\otimes_E\sV^*E\times\mathbb R\to\sT M\otimes_E\sV^*E$.

\subsection{Higher order jet bundles}\label{ssec:2.1.2}

Consider the bundle $\pi_1:\sJ E\to M$. The bundle of first jets of sections of $\pi_1$ is by definition an affine bundle
$$(\pi_1)_{1,0}:\sJ\sJ E\to \sJ E, \quad  \sj_m\psi\longmapsto\psi(m),    $$
where $\sj_m\psi$ is the first jet of a section $\psi:M\to\sJ E$ at the point $m$. It turns out that the projection 
$$\sj\pi_{1,0}:\sJ\sJ E\to\sJ E, \quad  \sj_m\psi\longmapsto\sj_m(\pi_{1,0}\circ\psi)$$
is an affine bundle as well. The bundle $\sJ\sJ E$ has the structure of a double affine bundle \cite{GRU} represented by the diagram
$$\xymatrix{
  &  \sJ\sJ E  \ar[dr]^{(\pi_1)_{1,0}}  \ar[dl]_{\sj\pi_{1,0}} &  \\
  \sJ E   \ar[dr]^{\pi_{1,0}} & &  \sJ E   \ar[dl]_{\pi_{1,0}}  \\
& E&
}
$$
In the bundle $\sJ\sJ E$ we have a subbundle $\widehat{\J2 E}$ consisting of elements having the same projection on both sides, i.e.
$$ \widehat{\J2 E}:=\{ \sj_m\psi\in\sJ\sJ E, \quad \sj\pi_{1,0}(\sj_m\psi)=(\pi_1)_{1,0}(\sj_m\psi) \}. $$
The bundle $\widehat{\J2 E}\to \sJ E$ is called the bundle of {\it semiholonomic} jets \cite{Saund}. It is an affine bundle modelled on the vector bundle $\otimes^2\sT^*M\otimes_{\sJ E}\sV E\to \sJ E$. If $(q^i,u\ind{^\alpha},u\ind{^\beta_j}, z\ind{^\gamma_k}, w\ind{^\delta_l_m})$ are coordinates in $\sJ\sJ E$ then the subbundle $\widehat{\sJ E}$ is given by the condition $u\ind{^\beta_j}=z\ind{^\beta_j}$. Therefore, we have natural coodinates $(q^i,u\ind{^\alpha},u\ind{^\beta_j}, u\ind{^\beta_j_k})$ in $\widehat{\sJ E}$ induced from $\sJ\sJ E$. 

In $\widehat{\J2 E}$ there exists a subbundle $\J2 E\to\sJ E$, such that
$$ \J2 E:=\{ \sj_m\psi\in\sJ\sJ E, \quad \psi=\sj\phi \}. $$
The bundle $\J2 E\to \sJ E$ is called the bundle of {\it second jets}. The elements of $\J2 E$ are called second jets of the section $\phi$ or holonomic jets. The bundle $\J2 E\to\sJ E$ is an affine bundle modelled on the vector bundle $\vee^2\sT^*M\otimes_{\sJ E}\sV E\to \sJ E$, where $\vee^2\sT^*M$ is the subbundle of symmetric tensors in $\sT^*M\otimes_M\sT^*M$. In $\J2 E$ we have induced coordinates $(q^i,u\ind{^\alpha},u\ind{^\beta_j}, u\ind{^\beta_j_k})$, such that $u\ind{^\beta_j_k}=u\ind{^\beta_k_j}$. There are natural inclusions
$$ \J2 E \subset \widehat{\J2 E}\subset \sJ\sJ E.  $$
The structure of the bundle $\widehat{\J2 E}$ will play a particularly important role in the following sections. 

Let us mention that the bundle of second jets may be defined independently from its embedding in $\sJ\sJ E$. 
We say that two sections $\phi,\phi':M\to E$ are in relation $\sim$ if and only if
%\begin{equation}\label{drugiedżety}
%\phi\sim\phi' \quad \Longleftrightarrow\quad \phi(m)=\phi'(m), \quad  \frac{\d^2}{\d t^2}_{|t=0}(\phi\circ\gamma)= \frac{\d^2}{\d t^2}_{|t=0}(\phi'\circ\gamma) 
%\end{equation}
%$\phi\sim\phi'$
\begin{eqnarray}\label{drugiedżety}
\phi(m)&=&\phi'(m), \\
\frac{\d}{\d t}_{|t=0}(\phi\circ\gamma)&=& \frac{\d}{\d t}_{|t=0}(\phi'\circ\gamma),  \\
\frac{\d^2}{\d t^2}_{|t=0}(\phi\circ\gamma)&=& \frac{\d^2}{\d t^2}_{|t=0}(\phi'\circ\gamma),
\end{eqnarray}
where $\gamma:\mathbb R\to M$ is any smooth curve in $M$ such that $\gamma(0)=m$. The second jet of a section $\phi$ at the point $m\in M$ is the equivalence class of sections with respect to the above relation and we denote it by $\mathsf j^2_m \phi$. The total space $\J2 E$ is by definition a set of equivalence classes $\mathsf j^2_m \phi$ at each point in $M$. Similarly, extending the condition of equality of derivatives up to the $k$-th order, we can define the bundle of $k$-jets denoted by $\mathsf J^kE$.

\subsection{Principal bundles and adjoint bundles}

In this and subsequent subsection we will briefly recall the fundamentals of principal bundles with a particular focus on the notion of a connection in this class of bundles. We will also fix the notation necessary for our subsequent work. The reader is referred, e.g. to \cite{KN,EGH} for the standard exposition of the principal bundle geometry. Our introduction concerning this topic is in large based on \cite{JMFF}.

Let $G$ be a Lie group with the Lie algebra $\mg$. We denote by $P$ a smooth manifold such that $G$ acts on it from the right in a smooth, free and proper way. Then, the space $M:=P/G$ of orbits of the action of $G$ on $P$ is a smooth manifold as well. The bundle $\pi:P\to M$ is called a principal bundle with a structure group $G$. It is locally isomorphic to $M\times G$. Let $U_\alpha\subset M$ be an open subset in $M$. A local trivialisation of $P$ is a $G$-equivariant diffeomorphism 
$$  \Psi_\alpha:\pi^{-1}(U_\alpha)\to U_\alpha\times G, \qquad   \Psi_\alpha(p)=(\pi(p),g_\alpha(p)),$$ 
where $g_\alpha:\pi^{-1}(U_\alpha)\to G$ is a $G$-valued function associated with the map $\Psi_\alpha$. The equivariance condition means that $\Psi_\alpha(pg)=\Psi_\alpha(p)g$, which implies that $g_\alpha$ is also $G$-equivariant, says $g_\alpha(pg)=g_\alpha(p)g$. Notice that the function $g_\alpha$ uniquely defines a local trivialisation of $P$. The transition between trivialisations $g_\alpha$ and $g_\beta$ defined on $\pi^{-1}(U_\alpha\cap U_\beta)$ is realised by the function  
$$  g_{\alpha\beta}:U_\alpha\cap U_\beta\to G,   \qquad  g_{\alpha\beta}(\pi(p))=g_\alpha(p)g_\beta(p)^{-1}.   $$
%\subsection{Adjoint bundles}
Let $F$ be a smooth manifold and let $G$ act on $F$ from the left. We introduce the action of $G$ on product $P\times F$ given by
$$ g(p,f)=(pg,g^{-1}f). $$
We denote by $N:=(P\times F)/G$ the space of orbits of this action. The bundle
$$\xi: N\to M,\quad  [(p,f)]\to \pi(p), $$
where $[(p,f)]$ is the orbit of the element $(p,f)\in P\times F$, is called an {\it associated bundle} of $P$ with a typical fiber $F$. Notice that the above projection does not depend on the choice of a representative in $[(p,f)]$, therefore it is well-defined. For $F$ being a vector space, the associated bundle is a linear bundle over $M$.
%Let $s_{\alpha}:M\supset\mathcal U_\alpha\to P$ be a local section of $P$ and let $N\supset\mathcal O_{\alpha}:=\xi^{-1}(\mathcal U_{\alpha})$. Then for each orbit $y\in\mathcal O_{\alpha}$, where $\xi(y)=m$, there exist a unique element $\chi_s(y)\in F$ such that $(s(m),\chi_s(y))$ belongs to $y$. A local section of $\pi$ provides then a local trivialization of $N$
%$$N\supset\mathcal O_\alpha\to \mathcal U_\alpha\times F,\quad y\longmapsto (\xi(y),\chi_s(y)).$$  
The most important examples of associated bundles of $P$ in the context of our work are the bundles with fibers $F=\mg$ and $F=G$, i.e. $N=(P\times\mg)/G$ and $N=(P\times G)/G$. From now on, we will use the notation 
$$\ad(P):=(P\times\mg)/G \quad\qquad  \textrm{and} \quad\qquad \Ad(P):=(P\times G)/G.$$ 
The action of $G$ on $\mg$ and the action of $G$ on $G$ is given by the adjoint map, namely
$$\Ad:G\times\mg\to\mg,\quad (g,X)\longmapsto\Ad_{g}(X),$$ 
$$\Ad:G\times G\to G,\quad (g,h)\longmapsto\Ad_{g}(h),   $$
respectively. For the sake of the simplicity of our notation we  have denoted both actions by the same symbol $\Ad$.

%{\red Maybe here about forms with values in adP?}

Denote by $\Omega^k(M,\mg)$ the bundle of $\mg$-valued $k$-forms on $M$. Let $\{U_\alpha\}$ be an open covering of $M$ and let $\{\xi_\alpha\}$ be a family of local $k$-forms on $M$ such that $\xi_\alpha\in\Omega^k(U_\alpha,\mg)$ for each $\alpha\in I\subset\mathbb R$. We also require that for each overlapping $U_{\alpha}\cap U_{\beta}$ the condition
\begin{equation}\label{ad}
\xi_\alpha(m)=\Ad_{g_{\alpha\beta}(m)}\circ\xi_\beta(m),   \qquad  m\in U_{\alpha\beta},\quad  g_{\alpha\beta}:U_{\alpha\beta}\to G  
\end{equation}
is satisfied. We claim that the family of $k$-forms $\{\xi_\alpha \}$ defines a $k$-form on $M$ with values in $\ad P$.  The space of $\ad P$-valued $k$-forms on $M$ will be denoted by $\Omega^k(M,\ad P)$. %Indeed, if $(v_1,...,v_k)\in\sT_mM$ are tangent vectors at a point $m\in U_{\alpha\beta}$, then $\xi_\alpha(v_1,...,v_k)$ and $\xi_\beta(v_1,...,v_k)$ satisfy (\ref{diag}). 

\subsection{Connection in a principal bundle}

Let $\sV P$ be the vertical subbundle in $\sT P$, i.e. the subbundle of tangent vectors, which are tangent to the fibers of $\pi:P\to M$. A connection in $\pi:P\to M$ is a $G$-invariant distribution $\mathcal D$ in $\sT P$, which is complementary to $\sV P$ at each point $p\in P$. By definition we have
\begin{equation}\label{dystrybhryzont}
\sT_pP=\mathsf V_pP\oplus\mathcal D_p,\qquad  p\in P
\end{equation}
and
\begin{equation}\label{dystrybniezmiennicz}
\mathcal D_pg=\mathcal D_{pg},  \quad g\in G.
\end{equation}
The above definition is very elegant and general, however, when it comes to applications, it is more convenient to represent a connection in a different way. We will start with introducing some basic mathematical tools. Let $X$ be an element of $\mg$. The group action of $G$ on $P$ defines the vertical vector field $\sigma_X$ on $P$ associated with the element $X$, namely
$$\sigma_X(p):=\ddt p\exp(tX). $$
The field $\sigma_X$ is called the fundamental vector field corresponding to the element $X$. The fundamental vector field is equivariant in the sense that
$$\sigma_X(pg)=\sigma_{\Ad_{g^{-1}}(X)}(p).$$
A connection form in $P$ is a $G$-equivariant, $\mg$-valued one-form $\omega$
$$\omega:\sT P\to\mg, $$
such that $\omega(\sigma_X(p)) = X$ for each $p\in P$ and $X\in\mg$. 
The $G$-equivariance means that 
$$R^*_g\omega(p)=\Ad_{g^{-1}}\circ\omega(p) .$$
It is easy to check that the distribution $\mathcal D_p:=\ker\omega(p)$ defines a connection in $P$. Since the connection form is an identity on vertical vectors, the difference of two connections is a horizontal form. It follows that the space of connections is the space of sections of an affine subbundle $\mathcal A\subset\sT^*P\otimes\mg$ modelled on the vector bundle of $\mg$-valued, $G$-equivariant horizontal one-forms on $P$. It turns out, that the space of such horizontal forms may be identified with the space of sections of the bundle $\sT^*M\otimes\ad P\to M$.

The connection one-form may be equivalently defined by means of a family of $\mg$-valued one-forms on $M$. This approach is widely used by physicists working in classical field theory. 
Let $s_\alpha:M\supset U_\alpha\to P$ be a local section. A pull-back of the connection form $\omega$ defines a $\mg$-valued one-form $A_\alpha:=s_\alpha^*\omega$.
%\begin{equation}\label{gaugefield}
%A_\alpha(m):\sT_mM\to\mg,\quad v_m\longmapsto\omega_P(s_\alpha(m))(\sT s_\alpha(v_m)). 
%\end{equation}
%The form $A_\alpha$ is called a {\it gauge field}. 
In local coordinates it reads
$$A_\alpha(m)=A_\alpha^a(m)\otimes e_a=A\ind{^a_j}(m)\mathsf dq^j\otimes e_a  \quad m\in M,$$
where $\{e_a\}$ is a basis of $\mg$. In order to simplify the notation we have skipped the index $\alpha$ in $A\ind{^a_j}$. One can show that two one-forms $A_\alpha$ and $A_\beta$ must satisfy the gluing condition
%$$A_\alpha=\Ad_{g_{\alpha\beta}}\circ(A_\beta-g^*_{\alpha\beta}\theta)   $$
%or equivalently
$$  A_\alpha=\Ad_{g_{\alpha\beta}}\circ A_\beta+g^*_{\beta\alpha}\theta   $$
on the overlapping $U_{\alpha\beta}$. In the above formula we have used the Maurer-Cartan form
$$\theta:G\to\sT^*G\otimes\mg,\quad \theta(g)=\sT L_{g^{-1}}.   $$
On the other hand, once we have a family of forms $\{A_\alpha\}$ we can restore a connection form $\omega$ on $P$. The restriction of $\omega$ to $\pi^{-1}(U_\alpha)$ is given by
\begin{equation}\label{cut}
\omega_{\alpha}(p)=\Ad_{{g_\alpha(p)}^{-1}}\circ\pi^*A_\alpha(p)+g^*_\alpha\theta(p). 
\end{equation}
The proof of the above statement is rather technical so we will skip it here.

The curvature of the connection is the two-form $\Omega_\omega:=(\d\omega)^h$, where $(\d\omega)^h$ is the horizontal part of $\d\omega$. After some computations one can show that
\begin{equation}\label{krzywiznaOmega}
\Omega_\omega=\d\omega+\frac{1}{2}[\omega\wedge\omega],  
\end{equation}
where $[\omega\wedge\omega]$ is the bracket of forms on $P$ with values in $\mg$. The curvature form, same as the connection form, is equivariant in the sense that $R_g^*\Omega_\omega=\Ad_{g^{-1}}\circ\Omega_\omega$. The curvature form is horizontal and $G$-equivariant therefore it defines the $\ad P$-valued two-form 
\begin{equation}\label{Fomega}
 F_\omega:M\to\wedge^2\sT^*M\otimes_M\ad P.
\end{equation} 
%Okazuje się więc, że krzywizna koneksji $\omega$ może być równoważnie reprezentowana przez dwuformę $\Omega$ na $P$ o wartościach w $\mg$ jak i przez dwuformę $F$ na $M$ o wartościach w $\ad P$. 
From now on we will use the following notation
\begin{align}
\cV & := \sT^*M\otimes_M\ad P, \label{eq:0}\\
 \cF & := \wedge^2\sT^*M\otimes_M\ad P.
\label{eq:1}
\end{align}
Section \ref{Fomega} can be also constructed from the local picture of $\Omega$. One can use $s_\alpha$ to pull-back the curvature form and obtain a $\mg$-valued two-form 
$$F_\alpha:=s^*_\alpha\Omega, \qquad F_\alpha=dA_\alpha+\frac{1}{2}[A_\alpha\wedge A_\alpha].$$ 
It is an easy exercise to check that each $F_\alpha$ and $F_\beta$ satisfy the condition (\ref{ad}) on the overlapping $U_{\alpha\beta}$. Therefore, we claim that the family of forms $\{F_\alpha \}$ define a section 

$$ F:M\to\wedge^2\sT^*M\otimes_M\ad P.$$  
It turns out that the curvature of the connection $\omega$ may be equally represented by the global $\mg$-valued two-form $\Omega$ on $P$ or by the global $\ad P$-valued two-form $F$ on $M$.

\subsection{Exterior covariant derivative}

Let $\Omega_G^k(P,V)$ be a subspace in $\Omega^k(P,V)$, which consists of $V$-valued $k$-forms that are both horizontal and equivariant. {\it The exterior covariant derivative} with respect to a connection $\omega$ and a representation $\rho: G\to GL(V)$ is the operator
$$D_\omega:\Omega_G^k(P,V)\to \Omega_G^{k+1}(P,V), \quad \zeta \longmapsto (\d\zeta)^h.  $$
One can show that %Można pokazać, że zachodzi wzór
\begin{equation}\label{pochodkowardef}
D_\omega\zeta=\d\zeta+\rho'(\omega)\wedge\zeta,   
\end{equation}
where $\rho'$ is given by
\begin{equation}\label{roprimdef}
\rho': \mg\to gl(V), \quad \rho'(X)v=\ddt\rho(\exp(tX))v
\end{equation}
and $\rho'(\omega)\wedge\zeta$ reads %oznacza jednocześnie iloczyn zewnętrzny form oraz działanie $\rho'(\omega)$ na $\zeta$
\begin{equation}\label{rhoprim}
 \rho'(\omega)\wedge\zeta(v_1,..,v_{1+k}):=\frac{1}{k!}\sum_{\sigma\in S_{1+k}} sgn(\sigma) \rho'\Big(\omega(v_{\sigma(1)} \Big)\zeta(v_{\sigma(2)},..,v_{\sigma(1+k)}). 
\end{equation}
Introducing a basis $(e_i)$ in $V$ we can write 
$$\omega=\omega^i\otimes e_i,\quad  \zeta=\zeta^j\otimes e_j,$$
$$\rho'(\omega)\wedge\zeta=\omega^i\wedge\zeta^j\otimes \rho(e_i)e_j.$$ 
Since $\Omega_G^k(P,V)\simeq \Omega^k(M, P\times_GV)$ %and $ \Omega_G^{k+1}(P,V)\simeq \Omega^{k+1}(M, P\times_GV)$ 
we have that the exterior covariant derivative in $\Omega_G^k(P,V)$ defines also an exterior covariant derivative in $\Omega^k(M, P\times_GV)$.

Consider forms on $P$ with values in the dual space $V^*$. The representation $\rho$ on $V$ defines the contragredient representation $\rho^\#$ on $V^*$. By definition we have 
$$\rho^\#: G\to End(V^*), \quad \rho^\#(g):={\rho(g)^*}^{(-1)}=\rho(g^{-1})^*.   $$
%gdzie $*$ oznacza odwzorowanie sprzężone. Dla macierzy $^*$ jest po prostu transpozycją. 
One can check that the action of $\mg$ on $V^*$ reads
$$ \rho^\#: \mg\to End(V^*), \quad {\rho^\#}'(X):=-\rho '(X)^*.  $$
Therefore, the covariant derivative $D_\omega$ in $\Omega_G^k(P,V)$ defines also a covariant derivative, say $D^\#_\omega$, in $\Omega_G^k(P,V^*)$ and $\Omega^k(M, P\times_GV^*)$. The total space $P\times_GV^*$ is the space of orbits of the action
$$G:P\times V^*\to P\times V^*,\quad g(p,f)=(pg,\rho(g^{-1})^\#f).  $$
For $\xi\in\Omega_G^k(P,V^*)$ we have
$$D^\#_\omega:\Omega_G^k(P,V^*)\to \Omega_G^{k+1}(P,V^*), \quad \xi \longmapsto (\d\xi)^h,  $$
$$D^\#_\omega\xi= \d\xi+{\rho^\#}'(\omega)\wedge \xi=\d\xi-\rho'(\omega)^*\wedge\xi.   $$
In the end lets us recall that the space $\Omega(P,\mg)$ is equipped with the canonical graded bracket  
%Przestrzeń $\Omega(P,\mg)$ dodatkowo posiada naturalną strukturę algebry z gradacją. Gradowany nawias na $\Omega(P,\mg)$ zadany jest przez
\begin{equation}\label{nawiasgradowany}
[\cdot\wedge\cdot]:\Omega^k(P,\mg)\times\Omega^l(P,\mg)\to\Omega^{k+l}(P,\mg), \quad  (\alpha,\beta)\to [\alpha\wedge\beta], 
\end{equation}
such that
$$ [\alpha\wedge\beta](v_1,..,v_{k+l}):=\frac{1}{k!l!}\sum_{\sigma\in S_{k+l}} sgn(\sigma)[\alpha(v_{\sigma(1)},..,v_{\sigma(k)}), \beta(v_{\sigma(k+1)},..,v_{\sigma(k+l)})],  $$
where $[\cdot,\cdot]$ is the Lie bracket in $\mg$. %The abve bracket is a  Nawias (\ref{nawiasgradowany}) jest odwzorowaniem dwuliniowym będącym uogólnieniem iloczynu zewnętrznego na formy o wartościach w algebrze Liego.

\section{Tulczyjew triple in mechanics and field theory} 

\subsection{Tulczyjew triple in mechanics}\label{ssec:1.3}
We will present now a point of view on describing a mechanical system, which is alternative to the one present in most textbooks \cite{K}. Its essence lies in the so-called {\it Tulczyjew triple}. Tulczyjew triple enables us to describe systems in both Lagrangian and Hamiltonian approach and shows relation between the two. It is important to notice, that the triple is based only on {\it canonical} structures of proper bundles. We will not derive here the whole formalism whose origin lies in the reinterpretation of variational description of statical systems. One can find the thorough analysis with all the details in numeorus Tulczyjew papers e.g. \cite{WMT1,WMT2,WMT3,WMT4,WMT5,WMT6}.

The Tulczyjew triple is a geometrical structure presented in the followig diagram

\begin{equation}\label{t:9}
\xymatrix{
& & & \color{red}{D} \ar@{^{(}->}[d]&& \\
& {\mathsf T}^*{\mathsf T}^*M \ar[rd]^{\pi_{\mathsf T^*M}}& & {\mathsf T}{\mathsf T}^*M\ar[ll]_{\beta_M}\ar[rr]^{\alpha_M}\ar[ld]_{\tau_{\mathsf T^*M}}\ar[rd]^{\sT\pi_M} & &{\mathsf T}^*{\mathsf T}M \ar[ld]_{\pi_{\mathsf TM}} \\
 & &{\mathsf T}^*M \ar[rd] \ar@/_1pc/[ur]_{{\color{red}X_H}}  \ar@/^1pc/[ul]^{{\color{red}{\mathsf d}H}}& & {\mathsf T}M \ar[ld] \ar@/_1pc/[ur]_{\color{red}{{\mathsf d}L}}&&  \\
&&&M&&&
}
\end{equation}
The right-hand side of the diagram is related to Lagrangian formalism while the left-hand side to Hamiltonian one. Both formalisms are based on the same scheme and the only difference is in generating objects on both sides. We will discuss now more precisely Lagrangian and Hamiltonian description in the language of the Tulczyjew triple. For simplicity, we will consider only autonomous systems i.e. with no external forces.

Let $M$ be the configuration manifold of the system and $L:\sT M\longrightarrow \mathbb{R}$ its Lagrangian. {\it The dynamics} of the system is a subset
\begin{equation}\label{t:2}
\mathcal D:=\alpha_M^{-1}\circ\mathrm dL(\sT M)
\end{equation}
of $\sT\sT^*M$. Since $\mathsf dL(\sT M)$ is a Lagrangian submanifold in $\sT^*\sT M$, and $\alpha_M$ is a symplectomorphism, the dynamics is a Lagrangian submanifold in $\sT\sT^*M$. From the physical point of view it is a (possibly implicit) first order differential equation for a trajectory in the phase space. A curve \mbox{$\eta: \mathbb{R}\supset I\rightarrow\sT^\ast M$} is a phase trajectory if $\dot\eta(t)\in \mathcal{D}$ for $t\in I$.

Now let us assume that Hamiltonian $H:\sT^\ast M\rightarrow\mathbb{R}$ of the system does exist. The dynamics of the system is then the image of the Hamiltonian vector field $X_H$ i.e.
\begin{equation}\label{t:21}
\mathcal{D}=X_H(\sT^\ast M)=\beta_M^{-1}(\mathrm{d}H(\sT^\ast M)).
\end{equation}
Phase space trajectories are integral curves of the field $X_H$.

The dynamics of the system may be projected on $\sT M\times \sT^*M$. The projection $\Lambda=\sT\pi_{M}\times \tau_{\sT^\ast M}(\mathcal{D})$ is a subset of Cartesian product of $\sT M$ and $\sT^*M$ therefore it can be understood as a relation between these two manifolds. If the dynamics comes from Lagrangian it is a graph of the Legendre map
$\lambda=\zeta_M\circ\mathrm{d}L$. If the dynamics comes from Hamiltonian it is a graph of a map `in the opposite direction'. In general, it can be a relation that is not a map at all. In such a case in place of Lagrangian and Hamiltonian we have to consider more general objects generating Lagrangian submanifolds than functions on  manifolds \cite{LM,WMT7}.

\subsection{Tulczyjew triple in field theory}

The Tulczyjew triple for mechanics may be generalised to the classical field theory. The very general Tulczyjew triple for the first order field theory has already been derived in \cite{KG}. We will briefly present results obtained in this paper.

Let $\mathsf VE\to E$ be the bundle of vectors vertical with respect to the projection $\pi:E\to M$, and $\mathsf V^*E\to E$ its dual bundle. Again, fields are represented by sections of the fibre bundle $\pi: E\to M$ and the space of infinitesimal configuration is $\mathsf J^1E$. As a natural consequence of the variational calculus we obtain that the phase space of the system is the space $\mathcal P:=\mathsf V^*E\otimes\Omega^{m-1}$. The bundle $\mathcal P\to E$ is a vector bundle. Let us introduce coordinates $(q^i,y^a,p^b_j)$ in $\mathcal P$. The field analogue of the Tulczyjew triple is a diagram
$$\xymatrix{
\mathsf P\sJd E  & \mathsf J^1\mathcal P \ar[l]_{\quad \beta}  \ar[r]^{\alpha \quad \quad}  & \mathsf V^*\mathsf J^1E\otimes\Omega^m}.
$$
The bundle  $\mathsf P\sJd E$ is the so-called `affine phase bundle', which is an affine analog of the cotangent bundle \cite{AVgeom}. Again, the left-hand side of the diagram refers to the Hamiltonian description of the sytem, while the right-hand side to the Lagrangian one. The main elements in the triple are maps $\alpha$ and $\beta$ which constitute the Lagrangian and Hamiltonian description, respectively. Both maps are morphisms of double vector-affine bundles \cite{KU,GRU}. %{\red napisać o alfa i beta tu}

Let $L:\sJ E\to\Omega^m$ be the Lagrangian of a system. We denote by $\d^{\mathsf v}L$ the vertical differential of $L$, i.e. the restricton of $\d L$ to $\sV\sJ E\subset\sT\sJ E$. The phase dynamics of the system is a subset
\begin{equation}\label{d2}
\mathsf J^1\mathcal P\supset\mathcal D:=\alpha^{-1}\circ\mathrm dL(\mathsf J^1E).
\end{equation}
From the mathematical pont of view the dynamics is a first-order partial differential equation for phase trajectories in $\mathcal P$. We say that a section $p:M\to\mathcal P$ is a solution of the Lagrange field equations if %$j^1_xp\in\mathcal D$
%Mówimy, że cięcie $p:M\to\mathcal P$ jest rozwiązaniem polowych równań Lagrange'a jeśli
$$ \sj_mp\in\mathcal D.  $$
Let us introduce coordinates $(q^i,y^a, p^b_j,y^c_k, p^l_{dm})$ in $\mathsf J^1\mathcal P$. Then, the Lagrange field equations read%We say that the section $p:M\to\mathcal P$ is a solution of the Lagrange field equations if $j^1_xp\in\mathcal D$. 
%In coordinates we have
$$
\displaystyle \sum_j\frac{\partial p^j_b}{\partial q^j}= \frac{\partial L}{\partial y^b}, \qquad
\displaystyle p^j_b=\frac{\partial L}{\partial y^b_j}.
$$
The Euler-Lagrange equations for field theory are consequence of the Lagrange field equations. Let us also mention that, same as in the case of mechanics, we do not require any regularity of the Lagrangian and the definition of the momenta comes directly from the theory.

On the other hand, a Hamiltonian is not a function and not even a map with values in forms, but a section of the bundle $\sJd E\to\mathcal P$, where $\sJd E$ is the affine dual of $\sJd E$. The affine differential $\mathsf dH$ can be viewed
as a map $\mathsf dH : \mathcal P\to\mathsf P\sJd E$. It defines the phase dynamics $\mathcal D =\beta^{-1}(\mathsf dH(\mathcal P))$. All the rigorous results concerning a very general Tulczyjew triple for field theory may be found in \cite{KG}.

Let us briefly recall the construction of the maps $\alpha$ and $\beta$.
We will start with the map $\alpha$. We will use the symbol ${\bf t}:=\ddt$ to denote the operation of taking a tangent vector to a curve in $t=0$, and the functor $\sV$ will always mean verticality with respect to the projection on $M$. Let $\sJ\sV E$ be the first jet bundle of $\sV E\to M$ and $\sV\sJ E$ the subbundle of vertical vectors in $\sT\sJ E$. Each element of $\sJ\sV E$ and $\sV\sJ E$ is represented by a map
$$ \chi:M\times\mathbb R\to  E   $$
preserving the projection on $M$. The elements of $\sV\sJ E$ and $\sJ\sV E$ have a form ${\bf t}\sj_m\chi$ and $\sj_m{\bf t}\chi$, respectively. There is a canonical isomorphism
\begin{equation}\label{kappa2}
 \kappa:\sV\sJ E\to \sJ\sV E, \quad \quad {\bf t}\sj_m\chi\longmapsto \sj_m{\bf t}\chi.
\end{equation}
The above map is a generalisation of the canonical involution $\kappa_M:\sT\sT M\to\sT\sT M$ known in mechanics. One can check that $\kappa$ is well-defined, i.e. it does not depend on the choice of the representative $\chi$.

The map $\alpha$ is a dual map to $\kappa$. The bundle $\sV^*\sJ E\to\sJ E$ is the dual bundle of $\sV\sJ E\to\sJ E$. It turns out that there exists also a canonical evaluation (over $\sJ E$) between $\sJ\mathcal P$ and $\sJ\sV E$ given by
\begin{equation}\label{EwaluacjaJPJE}
\sJ\mathcal P\times_{\sJ E}\sJ\sV E \to \Omega^m, \quad (\sj_mp,\sj_m\delta\sigma)\longmapsto \d\langle p,\delta\sigma \rangle.   
\end{equation}
We will denote the above evaluation by $\langle\!\langle \sj_mp,\sj_m\delta\sigma \rangle\!\rangle$. The map $\alpha$, which is a generalisation of the Tulczyjew isomorphism in mechanics \cite{WMT6,KG}, is a dual map to $\kappa$, understood as a morphism of the vector bundles $\sJ\sV E\to\sJ E $ and  $\sV\sJ E\to\sJ E$. The duality is taken with respect to the canonical evaluation between $\sV\sJ E$ and $\sV^*\sJ E\otimes_{\sJ E}\Omega^m$ and evaluation (\ref{EwaluacjaJPJE}). However, it is not a duality in the ordinary sense, which comes from the fact that (\ref{EwaluacjaJPJE}) is degenerated. By definition we have
$$ \alpha:\sJ\mathcal P\to \sV^*\sJ E\otimes_{\sJ E}\Omega^m, \quad  \langle \alpha(u), w\rangle =\langle\!\langle  u,\kappa(w) \rangle\!\rangle.   $$
In coordinates the map $\alpha$ has a form
\begin{equation}\label{alfa1}
\alpha(q^i,y^a,p\ind{^j_b}, y\ind{^c_k}, p\ind{^l_d_m})= (q^i,y^a,y\ind{^c_k}, \sum_l p\ind{^l_d_l}, p\ind{^j_b} ). 
\end{equation}
Notice that, in a contrast to mechanics, the map $\alpha$ is not an isomorphism of bundles.

On the Hamiltonian side we have a canonical diffeomorphism
\begin{equation}\label{Rdladżetów}
R_{\sJ E}:\sV^*\sJ E\otimes_M\Omega^m\to \mathsf P\sJd E. 
\end{equation}
We define $\beta$ as a composition
$$ \beta:=R_{\sJ E}\circ\alpha, \quad \beta:\sJ\mathcal P\to\mathsf P\sJd E.   $$
The map $\beta$ is a generalisation of the map $\beta_M$ in \ref{t:9} and it constitutes the basis of the Hamiltonian description of field theory. In coordinates we have 
\begin{equation}\label{betadladżetów}
 \beta(q^i,y^a,p\ind{^j_b},y\ind{^c_k},p\ind{^l_d_s})=(q^i,y^a,p\ind{^j_b},\sum p\ind{^k_c_k}, y\ind{^c_k}). 
\end{equation}
A Hamiltonian in classical field theory is not a function on the phase space with values in $\Omega^m$ but a family of sections of the affine bundle  %lecz rodziną cięć odpowiedniej wiązki afinicznej 
$$ \Sigma:\sJ E\times_E\mathcal P\to\sJd E   $$
parametrised by points in $\sJ E$. Notice that the fibers of the bundle $\sJd E\to\sJ E\times_E\mathcal P$ are one-dimensional. The above family of sections corresponds to the family of functions %Powyższa rodzina cięć odpowiada rodzinie funkcji
\begin{equation}\label{hamiltTP}
H:\sJ E\times_E\sJd E\to\Omega^m 
\end{equation}
with values in $\Omega^m$. In some cases the family of sections and the corresponding family of functions reduces to the single generating section and function
%W szczególnych sytuacjach rodzina cięć i odpowiadająca jej rodzi\-na funkcji o wartościach w $\Omega^m$ redukuje się do pojedynczego cięcia i odpowiadającej jej funkcji
$$ \Sigma: \mathcal P\to\sJd E,   $$
$$H: \sJd E\to\Omega^m. $$
%$H:\sJd E\to\Omega^m$. 
The dynamics is given then by the formula
$$\mathcal D=\beta^{-1}(\d H(\mathcal P)).   $$
%Zauważmy, że tak jak w przypadku strony lagranżowskiej proces generowania dynamiki jest bardzo prosty.

\section{Vector-affine products of bundles}

\subsection{Vector-affine product}\label{sec:1iloczyny}

We will discuss now %the problem of 
a particular case of an affine bundle, which can be decomposed on a properly defined sum of its affine subbundle and certain vector bundle. Before we will move to fibrations let us analyse this problem in a simpler case. Let $A$ be an affine space modelled on a vector space $V$ and let $B\subset A$ be an affine subspace in $A$ modelled on a vector subspace $V_1\subset V$. Moreover, we will assume that there exists a subspace $V_2\subset V$ such that $V$ is a direct sum $V=V_1\oplus V_2$. Let us choose an element $b_0\in B$. Then, any element $a\in A$ can be written in the form
$$a=b_0+v=b_0+v_1+v_2,$$
where the decomposition $v=v_1+v_2$ on $v_1\in V_1$ and $v_2\in V_2$ is unique. We introduce the element $b=b_0+v_1$, so that $a=b+v_2$. Notice that $b$ depends only on $a$, but it does not depend on $b_0$. Indeed, if $b'_0=b_0+u$ for a certain $u\in V_1$ and
$$ a=b'_0+v'=b'_0+v'_1+v'_2=b'+v'_2 $$
then $v=u+v'$. From the uniqueness of the decomposition $V=V_1\oplus V_2$ we obtain that $v_1=u+v'_1$ and $v_2=v'_2$, so that
$$b'=b'_0+v'_1=b_0+u+v_1-u=b_0+v_1=b.  $$
Therefore, the map
$$ A\to B\times V_2, \quad a\longmapsto (b,v_2) $$
is well-defined. Above decomposition will be called a vector-affine sum of $B$ and $V_2$ and we will denote it by $A=B\bar\oplus V_2$.

The above construction may be generalised to affine bundles over a manifold. Let  $A\to N$ be an affine bundle modeled on a vector bundle $V\to N$, and let $V$ be a Whitney sum of vector bundles $V_1\to N$ and $V_2\to N$. If $B\subset A$ is an affine subbundle in $A$ modelled on a vector subbundle $V_1$, then we have a decomposition
\begin{equation}\label{decomposition1}
A=B\bar\oplus_NV_2
\end{equation}
over $N$. By definition, each fiber of $B\bar\oplus_NV_2\to N$ over a point $n\in N$ is a vector-affine sum $B_n\bar\oplus V_{2n}$, where $B_n$ and $V_{2n}$ are fibers of $B\to N$ and $V_2\to N$, respectively.

As an example of the above construction let us consider
\begin{eqnarray}
A&=&\widehat{\mathsf J^2E},\\
B&=&\J2 E, \\
V_1&=&\sJ E\times_E(\vee^2\sT^*M\otimes_M\sV E),\\
V_2&=&\sJ E\times_E(\wedge^2\sT^*M\otimes_M\sV E),\\ 
N&=&\sJ E,
\end{eqnarray}
where $\vee^2\sT^*M$ is the space of symmetric tensors on $\sT^*M$.
By means of (\ref{decomposition1}) we have
\begin{equation}\label{decomposition}
\widehat{\mathsf J^2E}=\J2 E\bar\oplus_{\sJ E}(\sJ E\times_E\wedge^2\sT^*M\otimes_M\sV E). 
\end{equation}
Informally, we can say that each semiholonomic jet consists of the second jet of a certain section of a bundle $E\to M$ and a suitable antisymmetric two-vector on $M$ with values in $\sV E$.

\subsection{Tangent and cotangent bundle of vector-affine product}\label{sec:reduksympl}

Let us consider the tangent and cotangent bundle of a vector-affine product of bundles. We will start with the tangent bundle. Let $A\to N$ be an affine bundle modelled on a vector bundle $V\to N$. The tangent bundle $\sT A$ is a double affine-vector bundle \cite{GRU, KU} represented by the diagram
$$\xymatrix{
 \sT A  \ar[d]_{\tau_A }  \ar[r]  & \sT N \ar[d]^{\tau_N} \\
 A  \ar[r] &  N     
}
$$ 
The bundle $\sT A\to\sT N$ is an affine bundle modeled on the vector bundle $\sT V\to\sT N$. The addition of elements $v\in\sT A$ and $w\in\sT V$ having the same projection on $\sT N$ is defined in a following way. Let $\gamma_v,\gamma_w$ be curves representing $v$ and $w$, respectively, and having the same projection on $N$. Then, we can construct a curve
$$(\gamma_v+\gamma_w)(t):=\gamma_v(t)+\gamma_w(t)$$ 
and define the element $v\dot +w$ as a tangent vector of the above curve in $t=0$. This vector is fixed in $\tau_A(v)+\tau_V(w)$ and it has the same projection on $\sT N$ as $v$ and $w$. The affine structure of $\sT A\to\sT N$ is therefore a result of the application of the tangent functor to the structure of $A\to N$. In a similar way this functor acts on the vector-affine sum. Let us notice, that each representative $\gamma_v:\mathbb R\to A$ of the tangent vector $v\in\sT_aA$ may be written in the form $\gamma_v(t)=\gamma_B(t)+\gamma_{V_2}(t)$, where $\gamma_B:\mathbb R\to B$ and $\gamma_{V_2}:\mathbb R\to V_2$ are unique curves determined by the decomposition $A=B\bar\oplus V_2$ and having the same projection on $N$. Therefore, we can write
$$ v=v_B\dot +w_2, \qquad$$ 
$$ v_B=\ddt\gamma_B \in\sT_bB, \quad w_2=\ddt\gamma_{V_2}\in\sT_{p_2}V_2, $$
where $a=(b,p_2)$. Notice that $v,v_B$ and $w_2$ have the same projection on $\sT N$. %, a dodawanie $\dot +$ odbywa się w ramach wiązki $\sT A$.  
Finally, we obtain an isomorphism $\sT A\simeq\sT B\oplus_{\sT N}\sT V_2$ and the diagram
$$\xymatrix{
 \sT B\bar\oplus_{\sT N}\sT V_2  \ar[d]_{}  \ar[r]  & \sT N \ar[d]^{\tau_N} \\
 B\bar\oplus_{N}V_2  \ar[r] &  N     
}
$$ 
Let us consider now the cotangent bundle $\sT^*A=\sT^*(B\bar\oplus_{N}V_2)$. Same as $\sT A$, it is a double affine-vector bundle represented by a diagram
$$\xymatrix{
 \sT^*A  \ar[d]_{\pi_A }  \ar[r]  & V^* \ar[d]^{pr_N} \\
 A  \ar[r] &  N     
}
$$ 
The bundle $\sT^*A\to V^*$ is an affine bundle, in which the projection is constructed in a similar way as the projection $\sT^*E\to E^*$ for $E\to N$ being a vector bundle \cite{GRU,KU}. In short, each covector $\varphi\in\sT_a^*A$ may be restricted to the subspace $\sV_aA\simeq V_x$, and this restriction is identified with an element of $V_x^*$.

Let us recall a general fact from linear algebra that if $V$ is a vector space and $W\subset V$ is a subspace then $W^*\simeq V^*/W^o$, where $W^o\in V^*$ is an annihilator of $W$ in $V$. It means that there exists a projection $V^*\to W^*$ but in general there is no canonical embedding of $W^*$ in $V^*$. On the level of fiber bundles this fact has the following implications. If $V_1\to N$ and $V_2\to N$ are vector bundles, then in an obvious way we have
$$ \sT(V_1\times V_2)\simeq\sT V_1\times\sT V_2, \qquad  \sT^*(V_1\times V_2)\simeq\sT^*V_1\times\sT^*V_2. $$
The bundle $V_1\oplus_NV_2$ is a vector subbundle in $V_1\times V_2$, which means that $\sT(V_1\oplus_NV_2)=\sT V_1\oplus_{\sT N}\sT V_2$ and there exists a projection
$$ \sT^*(V_1\times V_2)\rightarrowtriangle \sT^*(V_1\oplus_NV_2),   $$
which is a symplectic relation. This relation may be obtained by choosing a subbundle \\ \mbox{$\sT^*(V_1\times V_2)\supset\sT_{V_1\times_NV_2}^*(V_1\times V_2)$}, where $\sT_{V_1\times_NV_2}^*(V_1\times V_2)$ is the set of those covectors in $\sT^*(V_1\times V_2)$, which are fixed at $V_1\times_NV_2$, and subsequently dividing it by the annihilator $\sT(V_1\oplus_NV_2)^o$, so that
$$\sT^*(V_1\oplus_NV_2)\simeq\sT_{V_1\times_NV_2}^*(V_1\times V_2)/\sT(V_1\oplus_NV_2)^o.$$ 
A similar situation occurs in the case of vector-affine sum, i.e. when we replace $V_1$ by $B$. The corresponding relation reads
$$ \sT^*B\times\sT^*V_2\rightarrowtriangle \sT^*(B\bar\oplus_NV_2).   $$
The bundle $\sT^*(B\bar\oplus_NV_2)$ has the structure of a double affine-vector bundle represented by the diagram
$$\xymatrix{
 \sT^*(B\bar\oplus_NV_2)  \ar[d] \ar[r]  & V_1^*\oplus_NV_2^* \ar[d]  \\
 B\bar\oplus_{N}V_2  \ar[r] &  N     
}
$$ 
The bundle $ \sT^*(B\bar\oplus_NV_2) \to  V_1^*\oplus_NV_2^*$ is an affine bundle modelled on the vector bundle $\sT^*V\to V^*$. Since now we will denote it by
$$ \rho:\sT^*(B\bar\oplus_NV_2) \to  V_1^*\oplus_NV_2^*. $$
Let us stress, that in a sharp contrast to the tangent bundle $\sT(B\bar\oplus_NV_2)$, the bundle $\sT^*(B\bar\oplus_NV_2)$ is not a direct product of bundles. 

Let us consider now the set $\rho^{-1}(0\oplus_NV_2^*)\subset \sT^*(B\bar\oplus_NV_2) $. By definition it consists of differentials of functions $f:B\bar\oplus_NV_2^*\to\mathbb R$, which are constant in the direction of $B$. The set $\rho^{-1}(0\oplus_NV_2^*)$ may be subject to symplectic reduction
$$   \sT^*(B\bar\oplus_NV_2)\supset\rho^{-1}(0\oplus_NV_2^*)\to\sT^*V_2.   $$
The above reduction will play an important role in the Lagrangian description of gauge field theories.

\section{The bundle of connections}

\subsection{The bundle of connections}

Let $E\to M$ be a fiber bundle and let $\sj_m\phi\in\sJ E$ be the first jet of a section $\phi:M\to E$ at the point $m\in M$. The tangent space to $\phi(M)$ at the point $\phi(m)$ defines the decomposition of the tangent space $\sT_{\phi(m)}E$ on the direct sum of the vertical subspace and the subspace, which is tangent to the image of the section $\phi$, i.e. 
$$ \sT_{\phi(m)}E=\sV_{\phi(m)}E\oplus\sT\phi(\sT_mM). $$   
Let us notice that $\sT\phi(\sT_mM)$ does not depend on the choice of a representative in $\sj_m\phi$. By taking a collection of first jets at each point $e\in E$ we obtain the decomposition of each tangent space $\sT_eE$. We conclude that a connection in $E\to M$ may be defined as a smooth section $\Gamma:E\to\sJ E$.% of the bundle $\pi_{1,0}:\sJ E\to E$.

From the above considerations we obtain that a connection in a principal bundle $P$ can be given by a global section of the bundle $\sJ P\to P$. The connection has to be compatible with the group action of $G$. Notice that the right action $R_g$ of $G$ on $P$ may be lifted to the action on $\sJ P$ given by
$$\sJ R_g:\sJ P\to \sJ P: \quad  \sj_m\phi\to \sj_m(\phi g),  $$
where $\phi g$ is a section given by the formula $(\phi g)(m):=\phi(m)g$. From the above definition we obtain that the diagram
$$\xymatrix{
\sJ P   \ar[r]^{\sJ R_g}  \ar[d]_{\pi_{1,0}}  & \sJ P   \ar[d]^{\pi_{1,0}}  \\
P  \ar[r]^{R_g} &  P
} 
$$ 
is commutative. From now on we will say that a section $\Gamma:P\to\sJ P$ is invariant with respect to the action of $G$ if $\Gamma(pg)=\Gamma(p)g$ for each $p\in P, g\in G$. Principal connections in $P$ are in a one-to-one correspondence with invariant sections of $\sJ P\to P$. The invariance of $\Gamma$ is equivalent to the invariance of the horizontal distribution in $\sT P$ defined by $\Gamma$. Let us notice that each invariant section $\Gamma$ defines a unique section 
$$\omega:M\to C, \quad C:=\sJ P/G.  $$
The bundle $C\to M$ turns out to be the bundle of principal connections in $P$. Each section of $C$ represents a gauge field of a given theory. Since $\sJ P\to P$ is an affine bundle modelled on the vector bundle $\sT^*M\otimes_P\sV P\to P$, we obtain that $C\to M$ is an affine bundle as well and it is modelled on the vector bundle $\sT^*M\otimes_M\ad P\to M$.

\subsection{Local picture of the connection bundle}\label{podrozdz:5.2}

Let $g_\alpha:\pi^{-1}(U_\alpha)\to U_\alpha\times G$ be a local trivialisation of $P$. It induces the local trivialisation of $\sJ P$ given by
\begin{equation}\label{TrywJP}
\Psi_\alpha:\sJ P\supset\pi_1^{-1}(U_\alpha) \to  G\times (\sT^*M\otimes\mg),  \qquad  (\sj_m\phi)\longmapsto\Big(g_\alpha(\phi(m)),(g_\alpha\circ\phi)^*\theta\Big),
\end{equation}
where
\begin{equation}\label{faktjakistam}
(g_\alpha\circ\phi)^*\theta(m):=-\Ad_{{g_\alpha(p)}^{-1}}\circ A_\alpha(m),  \quad p=\phi(m).
\end{equation}
In the above formula we have denoted by $\theta$ the Maurer-Cartan form on $G$. For two different trivialisations $\Psi_\alpha$ and $\Psi_\beta$ we have the transition function
$$\Psi_\beta\circ \Psi_\alpha^{-1}:G\times\sT^*M\otimes\mg\to G\times\sT^*M\otimes\mg, $$
\begin{equation}\label{fi}
\Big( g, B(m) \Big)\longmapsto  \Big(  g_{\beta\alpha}(m)g, B(m)+\Ad_{g^{-1}}\circ g_{\beta\alpha}^*\theta(m)  \Big). 
\end{equation}
Using (\ref{TrywJP}) we can also write the action of the group $G$ on $\sJ P$, i.e. 
$$\sJ R_h: G\times (\sT^*M\otimes\mg) \to  G\times (\sT^*M\otimes\mg),      $$
$$(g, j_\alpha)\longmapsto (gh,\Ad_{h^{-1}}\circ j_\alpha), \qquad j_\alpha\in\sT^*M\otimes\mg,  $$
where $R_h$ is the right action of an element $h\in G$ on $P$. To simplify the notation we have denoted by the same symbol $\sJ R_h$ both the first jet prolongation of $R_h$ and its trivialised version. 

Each trivialisation of $C$ comes from the trivialisation of $\sJ P$ divided by the action of $G$. Let $\sj_m\phi$ be a first jet, which in trivialisation $g_\alpha$ reads
\begin{equation}\label{trywialphi}
\sj_m\phi=\Big(g_\alpha(p),-\Ad_{g_\alpha(p)^{-1}}\circ A_\alpha \Big), \quad p=\phi(m).
\end{equation}
In our notation we will represent the equivalence class $[\sj_m\phi]$ by its representative in the neutral element of $G$. For instance, the equivalence class of $\Big(g_\alpha(p),-\Ad_{g_\alpha(p)^{-1}}\circ A_\alpha \Big)$ will be represented by the element $(e,- A_\alpha )$. It implies that the trivialisation of $C$ reads
$$C \to  \sT^*M\otimes\mg,  \qquad [\sj_m\phi]\longmapsto  -A_\alpha. $$
From the above formula we can see that sections of the bundle $C$ locally look like gauge fields. The transition formulas for two different trivialisations of $C$ coming from $g_\alpha$ and $g_\beta$ are the same as for $A_\alpha$, i.e. 
$$\xymatrix{
       &C    \ar[dr]^{\Psi_\beta}   \ar[dl]_{\Psi_\alpha} &             &        & [\sj_m\phi]      \ar[dr]^{\Psi_\beta}   \ar[dl]_{\Psi_\alpha}   &         \\   
\sT^*M\otimes\mg   \ar[rr]^{\Psi_\alpha^{-1}\circ\Psi_\beta} & & \sT^*M\otimes\mg      &    (j_\alpha)   \ar[rr]^{\Psi_\alpha^{-1}\circ\Psi_\beta \quad\qquad} & & ( \Ad_{g_{\beta\alpha}(m)}\circ j_\alpha+g_{\alpha\beta}^*\theta  )      \\
} $$

\subsection{The configuration bundle for gauge theories}\label{podrozdz431}

The bundle of infinitesimal configurations for a gauge field theory described by $C\to M$ is the bundle $\sJ C$. It is an affine bundle over $C$ modelled on the vector bundle
$$ \sT^*M\otimes_M\sV C\simeq C\times_M\sT^*M\otimes_M\sT^*M\otimes_M\ad P. $$
In the above formula we have used the isomorphism $\sV C\simeq C\times_M\sT^*M\otimes_M\ad P$, which comes from the fact that $C\to M$ is an affine bundle modelled on the vector bundle $\sT^*M\otimes_M\ad P\to M$. Therefore, the model bundle of $\sJ C\to C$ can be written as
$$ C\times_M\otimes^2\sT^*M\otimes_M\ad P\to C. $$
%Zauważmy, że z faktu iż $C\to M$ jest wiązką afiniczną modelowaną na wiązce wektorowej $\sT^*M\otimes_M\ad P\to M$ wynika, że wiązka $\sJ C\to M$ również jest wiązką afiniczną modelowaną z kolei na 
The affine structure of $C\to M$ implies that $\sJ C\to M$ is also an affine bundle modelled on the vector bundle $\sJ(\sT^*M\otimes_M\ad P)\to M$. In the following we will analyse in detail the internal structure of $\sJ C$. Since by definition $C=\sJ P/G$ we expect that there exists some relation between $\sJ C$ and the space of semiholonomic jets $\widehat{\J2 P}$. In particular, we will show that there exists an isomorphism $\sJ C\simeq \widehat{\J2 P}/G$.  

Let us start with the brief recalling of the space $\widehat{\J2 E}$ (see Section \ref{ssec:2.1.2}) in the case $E=P$. The bundle $\widehat{\J2 P}\to\sJ P$ is an affine bundle modelled on the vector bundle ${\sJ P}\times_P\otimes^2\sT^*M\otimes_P\sV P\to \sJ P$. Since $\sV P\simeq P\times\mg$ we obtain a natural identification
$$ {\sJ P}\times_P\otimes^2\sT^*M\otimes_P\sV P  \simeq   {\sJ P}\times_M(\otimes^2\sT^*M\otimes\mg) .   $$
From (\ref{decomposition}) we know that the bundle of semiholonomic jets $\widehat{\mathsf J^2P}$ has a natural decomposition
$$\widehat{\mathsf J^2P}=\J2 P\bar\oplus_{\sJ P}(\sJ P\times_P\wedge^2\sT^*M\otimes_M\sV P).  $$ 
Dividing the above product by $G$ we obtain 
\begin{equation}\label{rozkl2}
\widehat{\mathsf J^2P}/G=\J2 P/G\bar\oplus_C(C\times_M\wedge^2\sT^*M\otimes_M\ad P). 
\end{equation}
The bundle $\widehat{\mathsf J^2P}/G\to C$ is an affine bundle modelled on the vector bundle $C\times_M\sT^*M\otimes_M\sT^*M\otimes_M\ad P$. The model bundle has a natural decomposition 
$$ C\times_M\sT^*M\otimes_M\sT^*M\otimes_M\ad P=\Big(C\times_M\vee^2\sT^*M\otimes_M\ad P\Big)\oplus_C\Big(C\times_M\wedge^2\sT^*M\otimes_M\ad P\Big),$$
where $\vee^2\sT^*M$ is the subspace of symmetric tensors in $\otimes^2\sT^*M$. Notice that $\J2 P/G\to C$ is modelled on $C\times_M\vee^2\sT^*M\otimes_M\ad P\to C$.

Our purpose now is to show that there exists isomorphism $\widehat{\mathsf J^2P}/G\simeq \sJ C$. This isomorphism together with the decomposition (\ref{decomposition}) will allow us to "extract" geometrically the curvature from the first jet of a connection. In order to do that we will prove first two preparatory lemmas. Let us recall that each section $\omega:M\to C$ uniquely defines an invariant section $\bar\omega:P\to\sJ P$. Notice, that each jet $\bar\omega(p)\in\sJ_pP$ is represented by the equivalence class of sections $\phi_{\bar\omega}:M\to P$, such that $\phi_{\bar\omega}(m)=p$. The composition $\bar\omega\circ\phi_{\bar\omega}$ defines a local section of the bundle $\sJ P\to M$ in a neighborhood of $m\in M$. We have the following lemma

\begin{lemma}\label{twierdz2}
Let $\bar\omega:P\to\sJ P$ be an invariant section and let $\phi_{\bar\omega}:M\to P$ be a representative of $\bar\omega(p)\in\sJ_pP$ at a point $m\in M$, i.e. $\bar\omega(p)=\sj_m\phi_{\bar\omega}$. Then, we have
$$\sj_m\Big( \bar\omega\circ\phi_{\bar\omega} \Big) \in\widehat{\mathsf J^2P}.$$ 
\end{lemma}
{\bf Proof:}\\
We will show first that $\sj_m\Big( \bar\omega\circ\phi_{\bar\omega} \Big)$ does not depend on the choice of a representative $\phi_{\bar\omega}$. The first jet $\sj_m\Big( \bar\omega\circ\phi_{\bar\omega} \Big)$ is uniquely defined by the tangent map $\sT( \bar\omega\circ\phi_{\bar\omega}):\sT_mM\to\sT_{\bar\omega(p)}\sJ P$. From the composition rule for tangent maps we have
$$\sT_m(\bar\omega\circ\phi_{\bar\omega})=\sT_p\bar\omega\circ\sT_m\phi_{\bar\omega}. $$
Since $\sT_m\phi_{\bar\omega}$ does not depend on the choice of a representative in $\bar\omega(p)$, we have that the above map does not depend on the choice of $\phi_{\bar\omega}$ neither. Applying the projections $(\pi_1)_{1,0}$ and $\sj\pi_{1,0}$ to $\sj_m(\bar\omega\circ\phi_{\bar\omega})$ we obtain
$$(\pi_1)_{1,0}\Big(\sj_m(\bar\omega\circ\phi_{\bar\omega}) \Big)=(\bar\omega\circ\phi_{\bar\omega})(m)=\bar\omega(p) $$
$$\sj\pi_{1,0}\Big(\sj_m(\bar\omega\circ\phi_{\bar\omega})\Big)=\sj_m\Big(\pi_{1,0}\circ(\bar\omega\circ\phi_{\bar\omega})\Big)=
 \sj_m\phi_{\bar\omega}=\bar\omega(p), $$
which proves that indeed $\sj_m\Big( \bar\omega\circ\phi_{\bar\omega} \Big) \in\widehat{\mathsf J^2P}$.
$$\qquad\qquad\qquad\qquad\qquad\qquad\qquad\qquad\qquad\qquad\qquad\qquad\qquad\qquad\qquad\qquad \blacksquare$$
From now on we will denote by $[\sj_m\psi]$ the equivalence class of a first jet $\sj_m\psi\in\widehat{\mathsf J^2P}$ with respect to the group action of $G$. 

In the next step we will be interested in the following construction. Let us consider an invariant section $\bar\omega:P\to\sJ P$. By definition, at each point $p\in P$ we have $\bar\omega(p)=\sj_m\varphi$, where $\varphi:M\to P$ is a representative of the suitable equivalence class of sections. %W każdym punkcie $M$ możemy więc rozważać złożenie $\bar\omega\circ\varphi:M\to\sJ P$. 
Notice that the first jet of the section $\bar\omega\circ\varphi$ at the point $m=\pi(p)$ is independent on the choice of a representative $\varphi$ in $\sj_m\varphi$. In particular, for each representative $\varphi$ we have $\varphi(m)=p$. In the following it will be important to write down the first jet of $\bar\omega\circ\varphi$ in a given trivialisation $\sJ P\simeq G\times\sT^*M\otimes\mg$.
%W dalszych obliczeniach istotne będzie znalezienie postaci pierwszego dżetu cięcia $\bar\omega\circ\varphi$ w odpowiedniej trywializacji (przypomnijmy, że lokalnie $\sJ P\simeq G\times\sT^*M\otimes\mg$). 
Using Lemma 1 and the constructions from chapter \ref{podrozdz:5.2} we will prove a following lemma, which will simplify the calculations in the main theorem of this section. %rachunki w dalszej części tego rozdziału.
\begin{lemma}
Let
$$ \eta: M\to\sT^*M\otimes\mg,\quad  x\longmapsto  \Ad_{\varphi_\alpha(x)^{-1}} B_\alpha(x) $$
be a section of the bundle $\sT^*M\otimes\mg\to M$, where $B_\alpha:M\to\sT^*M\otimes\mg$ is a one-form with values in $\mg$ and $\varphi_\alpha: M\to G$ is a $G$-valued function, such that
$$A_\alpha(x):=-\Ad_{{\varphi_\alpha(x)}}\varphi_\alpha^*\theta(x)$$
(see \ref{faktjakistam}) and $B_\alpha(m)=A_\alpha(m)$ in a given point $m\in M$.
Then, the first jet of the section $\eta$ at the point $m\in M$ may be written as
$$ \sj_m\eta= \Big( \Ad_{\varphi_\alpha(m)^{-1}}(\sj_mB_\alpha) + \Ad_{\varphi_\alpha(m)^{-1}}\frac{1}{2}[B_\alpha(m)\wedge B_\alpha(m)] \Big)  \quad \in\sJ(\sT^*M\otimes\mg), $$
where
$$\Ad_{\varphi_\alpha(m)^{-1}}\sj_mB_\alpha := \sj_m\Big(\Ad_{\varphi_\alpha(m)^{-1}} B_\alpha\Big).    $$
\end{lemma}

{\bf Proof:}\\
The first jet $\sj_m\eta$ is uniquely defined by the tangent map $\sT\eta$. Let $v\in\sT_mM$ be a tangent vector represented by a curve, say $m(t)$, and such that $m(0)=m$. Applying $\sT\eta$ to $v$ we obtain 
$$ \sT\eta(v)=\ddt\eta\circ\gamma_v(t)=  \ddt\Big( \Ad_{\varphi_\alpha(m(t))^{-1}} B_\alpha(m(t))\Big)= $$
\begin{equation}\label{Teetarown}
= \ddt \Ad_{\varphi_\alpha(m(t))^{-1}} B_\alpha(m)+ \Ad_{\varphi_\alpha(m)^{-1}}\ddt B_\alpha(m(t)).
\end{equation}
The last equality comes from the fact that
%Ostatnią równość można otrzymać zapisując
$$\Ad_{\varphi_\alpha(m(t))^{-1}} B_\alpha(m(t))= \varphi_\alpha(m(t))^{-1} B_\alpha(m(t))\varphi_\alpha(m(t)).$$ 
%Notice that for the case when $B_\alpha$ is a zero-form with values in $\mg$, i.e. for a curve $t\to B(t)\in\mg$, the last equality reduces to a standard formula  
%$$ \ddt\Big( \Ad_{\varphi_\alpha(m(t))^{-1}} B(t)\Big)= \ddt\Ad_{\varphi_\alpha(m(t))^{-1}} B(0)+ \Ad_{\varphi_\alpha(m)^{-1}} \dot B(0).   $$
%Let us come back to the general case described in the lemma, i.e. when $B_\alpha$ is a one-form. 
From the second element in (\ref{Teetarown}) we obtain
$$\Ad_{\varphi_\alpha(m)^{-1}}\ddt B_\alpha(m(t))=\Ad_{\varphi_\alpha(m)^{-1}}\sT B_\alpha(v),$$
which means that we can identify it with $\Ad_{\varphi_\alpha(m)^{-1}}\sj_mB_\alpha(v)$. Let us consider now the first element in (\ref{Teetarown}), which is a bit more complicated. We will introduce a following notation. The tangent vector $\ddt \varphi_\alpha(m(t))$ is by definition an element of $\sT G$, which in trivialisation $\sT G\simeq G\times\mg$ reads $(\varphi_\alpha(m),X)$, where
$$X:=\varphi_\alpha(m)^{-1}\ddt \varphi_\alpha(m(t))\in\mg.$$ 
Notice that using (\ref{trywialphi}) we can write $X$ as
$$X=-\Ad_{\varphi_\alpha(m)^{-1}} A_\alpha(v),$$
where the relation between $\varphi_\alpha$ and $A_\alpha$ is the same as in (\ref{faktjakistam}) (recall that $\varphi_\alpha=g_\alpha\circ\phi$). From the assumptions of the lemma we know that at the point $m$ we have $B_\alpha(m)=A_\alpha(m)$, so we can replace $A_\alpha(m)$ by $B_\alpha(m)$. Then, in a neighborhood of $t=0$ we have
$$\varphi_\alpha(m(t))\sim \varphi_\alpha(m)\exp(tX)= \varphi_\alpha(m)\exp\Big(-t\Ad_{\varphi_\alpha(m)^{-1}}B_\alpha(m)(v)\Big),    $$ 
%a nie 
%$$  \varphi_\alpha(m(t))=\varphi_\alpha(m)\exp\Big(t A_\alpha(v_M))   \Big)  ?? $$}
%Wprowadźmy oznaczenie $X:=\Ad_{\varphi_\alpha(m)^{-1}}\circ A_\alpha(v_M)$. 
where $\varphi_\alpha(m(t))\sim \varphi_\alpha(m)\exp(tX)$ means that the tangent vectors to $\varphi_\alpha(m(t))$ and $\varphi_\alpha(m)\exp(tX)$ at $t=0$ are the same. Notice that there is a relation 
$$\Ad_{\varphi_\alpha(m(t))^{-1}}=\Ad_{\exp(-tX)}\Ad_{\varphi_\alpha(m)^{-1}},$$
where we have used the formula $(\exp(tX))^{-1}=\exp(-tX)$. 

Let us come back to the first element in (\ref{Teetarown}). Using the above conclusions we can write
$$\Big(\ddt \Ad_{\varphi_\alpha(m(t))^{-1}}\Big) B_\alpha(m)=\Big(\ddt \Ad_{\exp(-tX)}\Ad_{\varphi_\alpha(m)^{-1}}\Big) B_\alpha(m)=  $$
$$= \Big(\ddt \Ad_{\exp(-tX)}\Big)\Ad_{\varphi_\alpha(m)^{-1}}B_\alpha(m) =[-X,\Ad_{\varphi_\alpha(m)^{-1}} B_\alpha(m)]=  $$
$$[\Ad_{\varphi_\alpha(m)^{-1}} B_\alpha(m)(v), \Ad_{\varphi_\alpha(m)^{-1}}B_\alpha(m)]= \Ad_{\varphi_\alpha(m)^{-1}}[B_\alpha(m)(v), B_\alpha(m)]= $$
$$= \Ad_{\varphi_\alpha(m)^{-1}}\frac{1}{2}[B_\alpha(m)\wedge B_\alpha(m)](v,\cdot). $$
where $[\cdot,\cdot]$ is the Lie bracket in $\mg$ and $[\cdot\wedge\cdot]$ is the bracket %(\ref{nawiasgradowany}) 
of forms on $M$ with values in $\mg$. 
Using the identification $\sT_m\eta\sim\sj_m\eta$ and the calculations associated with (\ref{Teetarown}) we can finally write
$$\sj_m\eta=\Ad_{\varphi_\alpha(m)^{-1}} \sj_mB_\alpha+\Ad_{\varphi_\alpha(m)^{-1}}\frac{1}{2}[B_\alpha(m)\wedge B_\alpha(m)]. 
$$
Notice that $\sj_mB_\alpha\in\sJ(\sT^*M\otimes\mg)$ and $[B_\alpha(m)\wedge B_\alpha(m)]\in\sT^*M\otimes\sT^*M\otimes\mg$, which means that the above addition is well-defined. $\blacksquare$

Using the above lemmas we will prove now the crucial theorem of this section.

\begin{theorem}
The map
$$ \gamma: \sJ C\to\widehat{\mathsf J^2P}/G, \quad \sj_m\omega\longmapsto [\sj_m\Big( \bar\omega\circ\phi_{\bar\omega} \Big)]   $$
is an isomorphism of affine bundles over $C$ and its linear part is the identity on $\sT^*M\otimes_M\sT^*M\otimes_M\ad P$.% i.e. it is an isomorphism of bundles over $C$.
\end{theorem}
{\bf Proof:}\\
We will start with the derivation of $\gamma$ in a trivialisation. Let $g_\alpha:P\to G$ be a local trivialisation of $P$, which induces trivialisations $\sJ P\simeq G\times\sT^*M\otimes\mg$, $C\simeq\sT^*M\otimes\mg$ and $\sJ C\simeq\sJ(\sT^*M\otimes\mg)$. Let $\sj_m\omega\in\sJ_mC$ be a first jet, which in the above trivialisation has a form $\sj_mB_\alpha$, where $B_\alpha$ is, by definition, a representative of the equivalence class of sections 
$$B_\alpha:M\to\sT^*M\otimes\mg,$$
such that the tangent maps at the point $m$ given by each representative are the same. A family of sections $B_\alpha$ defines an invariant section $\bar\omega:P\to\sJ P$, which in our trivialisation reads
$$\bar\omega:M\times G\to G\times\sT^*M\otimes\mg,  \quad   \bar\omega(m,g_\alpha(p))=(g_\alpha(p),\Ad_{g_\alpha(p)^{-1}} B_\alpha(m)).   $$
To simplify the notation we have denoted by the same symbol $\bar\omega$ a section of the bundle $\sJ P\to P$ and its trivialised form.

Let us choose now a point $p\in P$, such that $\pi(p)=m$. Let $\phi:M\to P$ be a representative of % of sections defining a first jet $\bar\omega(p)$, i.e. 
$\bar\omega(p)=\sj_m\phi$, where $\phi(m)=p$. In the trivialisation we obtain
$$\bar\omega(m,g_\alpha\circ\phi(m))=(\varphi_\alpha(m),\Ad_{\varphi_\alpha(m)^{-1}} B_\alpha(m)), $$
%{\blue --------------------------------------------- dotąd jest ok ------------------------------------------}
where $\varphi_\alpha:=g_\alpha\circ\phi$. Notice that $\bar\omega(p)=\sj_m\phi$ implies that $m\longmapsto\Ad_{\varphi_\alpha(m)^{-1}}B_\alpha(m)$ satisfies assumptions of Lemma 2. Therefore, the first jet $\sj_m(\bar\omega\circ\phi)$ has a form %  Korzystając z wniosków z lematu 1 możemy zapisać pierwszy dżet powyższego cięcia w postaci
$$ \sj_m(\bar\omega\circ\phi)\in (G\times\sT^*M\otimes\mg)\times_{\sT^*M\otimes\mg}\sJ(\sT^*M\otimes\mg),  $$
$$\sj_m(\bar\omega\circ\phi)=\Big(\varphi_\alpha(m), \Ad_{\varphi_\alpha(m)^{-1}} B_\alpha(m), \Ad_{\varphi_\alpha(m)^{-1}}(\sj_mB_\alpha) + \Ad_{\varphi_\alpha(m)^{-1}}\frac{1}{2}[B_\alpha\wedge B_\alpha](m)\Big).$$
%{\blue --------------------------------------------- dalej jest ok ------------------------------------------}
Since both projections on $\sJ P$ are the same we can write
$$ \sj_m(\bar\omega\circ\phi)\in G\times\sJ(\sT^*M\otimes\mg),  $$
$$\sj_m(\bar\omega\circ\phi)=\Big(\varphi_\alpha(m), \Ad_{\varphi_\alpha(m)^{-1}}(\sj_mB_\alpha) + \Ad_{\varphi_\alpha(m)^{-1}}\frac{1}{2}[B_\alpha\wedge B_\alpha](m)\Big).$$
Taking the equivalence class of the above jet with respect to the group action we obtain
$$[\sj_m(\bar\omega\circ\phi)]=\Big( \sj_mB_\alpha+ [B_\alpha, B_\alpha](m)  \Big)=\Big( \sj_mB_\alpha+ \frac{1}{2}[B_\alpha\wedge B_\alpha](m)  \Big) \in\sJ(\sT^*M\otimes\mg),  $$
where $[B_\alpha\wedge B_\alpha]$ is a product %(\ref{nawiasgradowany}) 
of $\mg$-valued forms on $M$. Let us notice, that the above result is independent on the choice of a representative $\phi$ and, as a consequence, on the previously chosen point $p$. Furthermore, it does not depend on the choice of a representative $\omega$ in $\sj_m\omega$. Indeed, if $\omega,\omega'$ are two representatives of $\sj_m\omega$, which in trivialisation read $B_\alpha$ and $B'_\alpha$, respectively, then by definition $B_\alpha(m)=B'_\alpha(m)$ and $\sj_mB_\alpha=\sj_mB'_\alpha$.   %, a w konsekwencji
%$$ \sj_mB'_\alpha+ \frac{1}{2}[B'_\alpha(m)\wedge B'_\alpha(m)] =\sj_mB_\alpha+ \frac{1}{2}[B_\alpha(m)\wedge B_\alpha(m)].  $$
Finally, the map $\gamma$ reads
$$\gamma: \sj_mB_\alpha\longmapsto  \sj_mB_\alpha+ \frac{1}{2}[B_\alpha(m)\wedge B_\alpha(m)].  $$
Using the above formula we can easily show that $\gamma$ is an affine map whose linear part is the identity. Let $\sj_mB_\alpha\in\sJ(\sT^*M\otimes\mg)$ and $\sj_mD_\alpha\in\sJ(\sT^*M\otimes\mg)$ be elements of $\sJ_eC$. We have
$$ \gamma(\sj_mB_\alpha)-\gamma(\sj_mD_\alpha)= \sj_mB_\alpha+ \frac{1}{2}[B_\alpha(m)\wedge B_\alpha(m)] -\sj_mD_\alpha- \frac{1}{2}[D_\alpha(m)\wedge D_\alpha(m)]=$$
$$= \sj_mB_\alpha-\sj_mD_\alpha, $$
where in the last equality we have used the fact that $B_\alpha(m)=D_\alpha(m)$. From the above formula one can see that the linear part of $\gamma$ is the identity on $\sT^*M\otimes_M\sT^*M\otimes_M\ad P$. $\blacksquare$ 

Using the above theorem and (\ref{rozkl2}) we obtain the decomposition
\begin{equation}\label{rozklJC}
\sJ C=\J2 P/G\bar\oplus_C(C\times_M\wedge^2\sT^*M\otimes_M\ad P),
\end{equation}
$$\sj_m\omega=\Big([\mathsf j^2_m\bar\omega], \omega(m),F_\omega(m)\Big) $$
over $C$. In the above formula we have denoted by $[\mathsf j^2_m\bar\omega]$ the equivalence class of the second jet of the section $\bar\omega$ with respect to the group action and $F_\omega(m)$ is the value of the curvature form (\ref{Fomega}) associated with $\sj_m\omega$. 

Notice that a similar result to Theorem 1 was briefly discussed by Sardanashvily in \cite{SG2}. In the formula (6.6) in his book he presents the decomposition 
$$  \sJ C= \J2 P/G\bar\oplus_C(\wedge^2\sT^*M\otimes_M\ad P), $$
$$ k_{\mu\lambda}^m=\frac{1}{2} ( k_{\mu\lambda}^m+k_{\lambda\mu}^m+  c^m_{nl}k_{\lambda}^nk_{\mu}^l  )  +  \frac{1}{2} ( k_{\mu\lambda}^m-k_{\lambda\mu}^m-  c^m_{nl}k_{\lambda}^nk_{\mu}^l  ).  $$
where $(x^\mu,k^m_\mu,k_{\mu\lambda}^m)$ are coordinates in $\sJ C$.
However, his result is not true. First of all, we have $c^m_{nl}=-c^m_{ln}$, which implies that the element $\frac{1}{2} ( k_{\mu\lambda}^m+k_{\lambda\mu}^m+  c^m_{nl}k_{\lambda}^nk_{\mu}^l  )$ is not symmetric with respect to $\mu\lambda$, therefore it can not be part of $\J2 P/G$. Secondly, \cite{SG2} does not contain any calculations that would support his statements.

\subsection{The phase bundle}

%The second crucial object of a given physical theory, apart from its configuration bundle, is the phase bundle. 
%In physics the structure of the phase bundle is usually derived from the analysis of variations of the action functional of a given theory.
%Postać wiązki fazowej w teoriach fizycznych z reguły wyprowadza się analizując wariacje działania $S$ dla danej teorii. 
In subsection 3.2 it was said that the phase bundle in field theory is represented by the bundle $\mathcal P=\sV^*E\otimes_E\Omega^{m-1}$, where $E\to M$ is the space of fields. From now on we will use the notation $\cV:=\sT^*M\otimes_M\ad P$. In gauge field theories the phase space is given by
$$\mathcal P=\sV^*C\otimes_C\Omega^{m-1}=C\times_M\cV^*\otimes_M\Omega^{m-1},$$
where $\cV^*$ is the dual bundle of $\cV$. Using isomorphism $\sT M\otimes_M\Omega^m\simeq\Omega^{m-1}$ we can write
$$\cV^*\otimes_M\Omega^{m-1}\simeq \sT M\otimes_M\ad^*P\otimes_M\sT M\otimes_M\Omega^{m}= \sT M\otimes_M\sT M\otimes_M\ad^*P\otimes_M\Omega^{m}. $$
Notice that the tensor bundle $\sT M\otimes_M\sT M$ has the canonical decomposition 
$$\sT M\otimes_M\sT M= \wedge^2\sT M\oplus_M\vee^2\sT M.$$
Using the above formula we obtain
$$ \sT M\otimes_M\sT M\otimes_M\ad^*P\otimes_M\Omega^{m}=(\wedge^2\sT M\oplus_M\vee^2\sT M)\otimes_M\Omega^m\otimes_M\ad^*P $$
$$= (\wedge^2\sT M\otimes_M\Omega^m\otimes_M\ad^*P)\oplus_M(\vee^2\sT M\otimes_M\Omega^m\otimes_M\ad^*P)=$$
\begin{equation}\label{wiazka1i2}
= (\Omega^{m-2}\otimes_M\ad^*P)\oplus_M(\vee^2\sT M\otimes_M\Omega^m\otimes_M\ad^*P).
\end{equation}
%{\blue Elementy $\sT M\otimes\sT M\otimes\ad^*P\otimes_M\Omega^{m}$ są postaci $X=X\ind{_a^i^j}\partial_i\otimes\partial_j\otimes e_*^a\otimes\eta$, gdzie $(e_*^a)$ są wektorami bazy dualnej do $(e_a)$.}
The phase bundle $\mathcal P\to C$ turns out to be the Whitney sum of vector bundles
\begin{eqnarray}
&&C\times_M(\Omega^{m-2}\otimes_M\ad^*P)\to C, \\
&&C\times_M(\vee^2\sT M\otimes_M\Omega^m\otimes\ad^*P)\to C.  
\end{eqnarray}
For the clarity of the presentation we will introduce the notation
$$\overline{\mathcal P}:=\Omega^{m-2}\otimes_M\ad^*P, $$ 
$$\mathcal S:=\vee^2\sT M\otimes_M\Omega^m\otimes\ad^*P. $$
The bundle $C\times_M\barP\to C$ will be called the {\it reduced phase bundle} and it represents the reduced phase space of the system. %Let us consider a bundle of first jets of the bundle $\mathcal P\to M$. 
Notice that (\ref{wiazka1i2}) implies that $\sJ\mathcal P$ is the product of bundles
\begin{equation}\label{rozklJP}
\sJ\mathcal P=\sJ C\times_M \sJ\barP\oplus_M\sJ\mathcal S.
\end{equation}
%{\blue 
%Mamy więc rozkłady
%$$\sT M\otimes_M\sT M\otimes_M\ad^*P\otimes_M\Omega^{m}= (\Omega^{m-2}\otimes_M\ad^*P)\oplus(\vee^2\sT M\otimes_M\Omega^m\otimes\ad^*P) $$
%}
%$$\sJ(\sT M\otimes_M\TM\otimes_M\ad^*P\otimes_M\Omega^{m})= \sJ(\Omega^{m-2}\otimes_M\ad^*P)\oplus\sJ(\vee^2\sT M\otimes_M\Omega^m\otimes_M\ad^*P)$$
The decompositions (\ref{rozklJC}) and (\ref{rozklJP}) will be crucial in the following part of our work. %our work later on. % będą kluczowe w dalszej części pracy.

\section{Lagrangian description}\label{sec:5.2}

In sections 5 we have carefully analysed the geometric structure of the bundle of connections. Now we will focus on the dynamics of gauge fields. The starting point for our considerations will be the Tulczyjew triple presented in section 3. In the previous section we have shown that the bundles $\sJ C$ and $\sJ\mathcal P$ have the natural structure of a product of bundles. In particular, we have derived the decomposition \ref{rozklJC} of the bundle $\sJ C$, which allows to project the first jet of a connection in a given point onto its value and curvature. As we mentioned before, in most of the physically interesting field theories Lagrangian does not depend on the entire first jet but only on some informations contained in it and associated with its antisymmetric part. Each description of the gauge fields dynamics should be therefore independent of the symmetry associated with the addition of the symmetric tensor to the first jet of a connection. In this section we will 
%use \ref{rozklJC} and \ref{rozklJP} to 
reduce the Tulczyjew formalism with respect to that symmetry.   % the symmetry described above.

\subsection{The structure of iterated bundles}\label{sec:5.1}

Let us consider the map $\kappa$ on a bundle $C$. To simplify the notation we will use the symbol ${\bf t}:=\ddt$ from the subsection 3.2. Notice that each element of $\sV\sJ C$ has a form ${\bf t}\sj_x\chi$, where $\chi$ is a map
\begin{equation}\label{hi}
\chi:\mathbb R\times M\to C      
\end{equation}
preserving the projection on $M$. By definition \ref{kappa2} the map $\kappa$ on $C$ reads
\begin{equation}\label{kappaC}
\kappa:\sV\sJ C\to\sJ\sV C,\quad {\bf t}\sj_x\chi\longmapsto \sj_x{\bf t}\chi.
\end{equation}
The affine structure of  the bundle $C$ allows to write the above equation in a simpler way. From the fact that $C\to M$ is an affine bundle modelled on the vector bundle $\cV= \sT^*M\otimes_M\ad P\to M$, we obtain that the bundle $\sJ C\to M$ is an affine bundle as well and it is modelled on the vector bundle $\sJ\cV\to M$. We have isomorphisms
\begin{align}
\sV C&\simeq C\times_M\cV,   \\
\sV\sJ C&\simeq\sJ C\times_M\sJ\cV, \label{eqn:1}\\    
\sJ\sV C&\simeq\sJ(C\times_M\cV)\simeq\sJ C\times_M\sJ\cV,   \label{eqn:2}
\end{align}
where for simplicity we have used the notation (\ref{eq:0}) and (\ref{eq:1}).

The map $\kappa$ can be expressed in terms of the above identifications. Since the bundle of connections is an affine bundle, each element ${\bf t}\sj_x\chi$ has the representative of the form
$$  \chi(t,x)=\omega(x)+t\tau(x), \qquad \omega:M\to C, \quad \tau:M\to\cV.  $$
By taking the first jet of the above map in $x\in M$ we obtain a vertical curve in $\sJ_xC$
$$\sj_x\chi(\cdot):\mathbb R\to\sJ C,  \quad   t\longmapsto \sj_x\omega+t\sj_x\tau.  $$
We can calculate the derivative of the above curve over $t$ in $t=0$ and obtain
$${\bf t}\sj_x\chi(t)=(\sj_x\omega,\sj_x\tau),  $$
where we have used the identfication (\ref{eqn:1}). If we now repeat the above procedure in the opposite order, i.e. we calculate the derivative of $\chi$ over $t$ in $t=0$ and then take the first jet, we will obtain
$$\sj_x{\bf t}\chi=(\sj_x\omega,\sj_x\tau),  $$
where we have used isomorphism (\ref{eqn:2}). In the above identifications the map $\kappa$ turns out to be the identity
$$\kappa:\sJ C\times_M\sJ\cV \to \sJ C\times_M\sJ\cV, \qquad (\sj_x\omega,\sj_x\tau)\longmapsto (\sj_x\omega,\sj_x\tau).   $$
Let us fix now an element $\sj_x\omega\in\sJ_xC$.  The first jet $\sj_x\omega$ has a canonical projection on the value $\omega(x)$, which defines a covariant derivative $D_\omega$ in $\cV_x$. Then, we have in $\sJ_x\cV$ a subspace 
$$S_x:=\{\sj_x\tau_0\in\sJ_x\cV, \quad (D_\omega\tau_0)(x)=0 \}.$$
Notice that each first jet $\sj_x\tau\in\sJ_x\cV$ can be decomposed on a pair $\Big(\sj_x\tau_0,D_\omega\tau(x)\Big)$, where $\tau_0:M\to\cV$ is a section of the bundle $\cV$, such that $\tau_0(x)=\tau(x)$ and $\sj_x\tau_0\in S_x$, and $D_\omega\tau(x)$ is the covariant derivative of the section $\tau$ in the point $x\in M$. Notice that $D_\omega\tau(x)$ does not depend on the choice of the representative in $\sj_x\tau$. Therefore, we obtain a decomposition 
\begin{equation}\label{rozkSF}
\sJ_x\cV= S_x\times\cF_x, \quad \sj_x\tau\longmapsto\Big(\sj_x\tau_0,D_\omega\tau(x)\Big).
\end{equation}
Using the above identification we can project $\sj_x\tau$ on $D_\omega\tau(x)$. Furthermore, there is the canonical projection $\sJ_x\cV\to\cV$ so that we obtain a map 
$$ \sJ\cV\to\cV\times_M\cF,  $$
\begin{equation}\label{rzutowanie1}
\sj_x\tau\longmapsto\Big(\tau(x), D_\omega\tau(x)\Big).
\end{equation}
Finally, the maps (\ref{rozklJC}) and (\ref{rzutowanie1}) define a projection
\begin{equation}\label{rzut12}
\sJ C\times_M\sJ\cV \to (C\times_M\cF)\times_M\cV\times_M\cF, \quad  (\sj_x\omega,\sj_x\tau)\longmapsto \Big(\omega(x), F_{\omega}(x), \tau(x), D_{\omega}\tau(x)\Big).
\end{equation}
Notice that isomorphisms (\ref{eqn:1}) and (\ref{eqn:2}) imply that the bundles $\sJ\sV C$ and $\sV\sJ C$ may be reduced by means of (\ref{rzut12}).
%{\red   Następnie, za pomocą odwzorowań (\ref{rozklJC}) i (\ref{rzutowanie1}) oraz identyfikacji (\ref{eqn:1}) i (\ref{eqn:2}) konstruujemy rzuty 
%\begin{equation}\label{rzut1}
%\sV\sJ C\to (C\times_M\cF)\times_M\cV\times_M\cF, \quad  {\bf t}\sj_m\chi\longmapsto \Big(\omega(m), F_{\omega}(m), \tau(m), D_{\omega}\tau(m)\Big), 
%\end{equation}
%\begin{equation}\label{rzut2}
%\sJ\sV C\to (C\times_M\cF)\times_M\cV\times_M\cF, \quad \sj_m{\bf t}\chi \longmapsto \Big(\omega(m), F_{\omega}(m), \tau(m), D_{\omega}\tau(m)\Big).
%\end{equation}   }
Therefore, we can use (\ref{rzut12}) to reduce the map $\kappa$ given by (\ref{kappaC}). The reduction of $\kappa$ is presented on the diagram 
$$\xymatrix{
 \sV\sJ C  \ar[d]^{(\ref{rzut12})}   \ar[r]^{\kappa} &  \sJ\sV C  \ar[d]^{(\ref{rzut12})}  \\
 (C\times_M\cF)\times_M\cV\times_M\cF  \ar[r]^{\overline\kappa}   & (C\times_M\cV)\times_M\cF\times_M\cF 
}
$$
and 
$$\xymatrix{
 (\sj_m\omega,\sj_m\tau)  \ar[d]^{(\ref{rzut12})}  \ar[r]^{\kappa} &  (\sj_m\omega,\sj_m\tau) \ar[d]^{(\ref{rzut12})}   \\
 \Big(\omega(x), F_{\omega}(x), \tau(x), D_{\omega}\tau(x)\Big)   \ar[r]^{\overline\kappa}  &  \Big(\omega(x), \tau(x), F_{\omega}(x), D_{\omega}\tau(x)\Big) 
}
$$
From the above diagram we see that the reduced map $\overline\kappa$ has a form
\begin{equation}\label{zredukkappa}
\overline\kappa:\Big(\omega(x), F_{\omega}(x), \tau(x), D_{\omega}\tau(x)\Big) \longmapsto \Big(\omega(x), \tau(x), F_{\omega}(x), D_{\omega}\tau(x)\Big).
\end{equation}

\subsection{The structure of the bundle $\sV^*\sJ C$}

Let $\sV^*\sJ C\to\sJ C$ be the dual bundle of $\sV\sJ C\to\sJ C$. Since $\sJ C\to M$ is an affine bundle modelled on the vector bundle $\sJ\cV\to M$ we have a canonical isomorphism $\sV^*\sJ C\simeq\sJ C\times_M\sJ{^*\cV}$, where $\sJ{^*\cV}\to M$ is the dual bundle of $\sJ\cV\to M$.

In section (\ref{sec:reduksympl}) we have considered an affine bundle $A\to N$ with an affine subbundle $B\to N$ modelled on a vector bundle $V_2\to N$. In particular we have obtained the fibration
$$ \rho:\sT^*(B\bar\oplus_NV_2) \to  V_1^*\oplus_NV_2^* $$
and the set $\rho^{-1}(0\oplus_NV_2^*)\subset \sT^*(B\bar\oplus_NV_2) $. By definition it consists of the differentials of functions $f:B\bar\oplus_NV_2\to\mathbb R$, which are constant in the direction of $B$. The set $\rho^{-1}(0\oplus_NV_2^*)$ is a coisotropic submanifold, which means that we can perform the symplectic reduction
$$   \sT^*(B\bar\oplus_NV_2)\supset\rho^{-1}(0\oplus_NV_2^*)\to\sT^*V_2.   $$
Let us apply the above construction to the case
\begin{eqnarray*}
A&=&\sJ_xC,\\
B&=&(\J2 P/G)_x, \\
V_1&=&C_x\times(\vee^2\sT_x^*M\otimes_M\ad_xP),\\
V_2&=&C_x\times\mathcal F_x,\\ 
N&=&C_x.  \\ %U&=&\Omega^m_x.  \\
\end{eqnarray*}
We introduce the notation
\begin{equation}\label{podwiazkaK}
\sV^*\sJ C\supset\mathcal K:=\rho^{-1}(0\oplus_{C}C\times_M\mathcal F^*).
\end{equation}
Using isomorphisms $\sT^*\sJ_xC\simeq\sV_x^*\sJ C$ and $\sT^*(C_x\times\mathcal F_x)\simeq\sV_x^*(C\times_M\mathcal F)$ we can perform the symplectic reduction 
$$ \sV_x^*\sJ C\supset\mathcal K_x\to\sV_x^*(C\times\mathcal F).   $$
If we take the above map point by point in $M$ we will obtain a reduction
\begin{equation}\label{podrozmaitK}
\sV^*\sJ C\supset\mathcal K\to\sV^*(C\times_M\mathcal F)\simeq  C\times_M\mathcal F\times_M\cV^*\times_M\mathcal F^*
\end{equation}
over $C\times_M\mathcal F$. In the following, we will be rather interested in the bundle $\sV^*\sJ C\otimes_{\sJ C}\Omega^m$ than $\sV^*\sJ C$ itself. The suitable reduction for $\sV^*\sJ C\otimes_{\sJ C}\Omega^m$ reads
\begin{equation}\label{podrozmaitK2}
\sV^*\sJ C\otimes_{\sJ C}\Omega^m\supset\mathcal K\to C\times_M\mathcal F\times_M\cV^*\otimes_M\Omega^m\times_M\mathcal F^*\otimes_M\Omega^m\simeq 
\end{equation}
$$ \simeq (C\times_M\cF)\times_M(\Omega^{m-1}\otimes_M\ad^*P)\times_M\barP. $$

\subsection{The Tulczyjew map for gauge theories}  

According to (\ref{alfa1}) the Tulczyjew map for gauge theories reads
$$\alpha:\sJ\mathcal P\to\sV^*\sJ C\otimes\Omega^m,   $$
$$(q^i,A\ind{^a_i}, X\ind{_a^i^k},A\ind{^a_i_j},X\ind{_a^i^k_s}) \longmapsto \Big(q^i,A\ind{^a_i},A\ind{^b_i_j}, \sum_k X\ind{_a^i^k_k}, X\ind{_a^i^k}\Big).  $$
Using the decomposition of $\sJ\mathcal P$ described in (\ref{rozklJP}) and the isomorphism $\sV^*\sJ C\simeq\sJ C\times_M\sJ{^*\cV}$, where $\sJ{^*\cV}\to M$ is a dual bundle of $\sJ\cV\to M$, we can write the Tulczyjew map in the form
\begin{equation}\label{alphadecomp}
\alpha: \sJ C\times_M\sJ\bar{\mathcal P}\times_M\sJ\mathcal S\to \sJ C\times_M\mathsf J^*\cV\otimes_M\Omega^m. 
\end{equation}
%$$(q^i,A\ind{^a_i},A\ind{^b_i_j})\times(q^i,X\ind{_a^i^k},X\ind{_a^i^k_s}) \longmapsto (q^i,A\ind{^a_i},A\ind{^b_i_j})\times\Big(q^i,\sum_k X\ind{_a^i^k_k},X\ind{_a^i^k}\Big),  $$
%where $\mathsf J^*\cV\to M$ is a dual bundle to $\sJ\cV\to M$. 
From (\ref{alphadecomp}) we can see that the structure of the Tulczyjew map is encoded in two maps 
$$\alpha_1:\sJ C\to \sJ C,   $$
being an identity, and
$$ \alpha_2:\sJ\bar{\mathcal P}\times_M\sJ\mathcal S\to  \mathsf J^*\cV\otimes_M\Omega^m.   $$ 
%$$ (q^i,X\ind{_a^i^k},X\ind{_a^i^k_s}) \longmapsto\Big(q^i,\sum_k X\ind{_a^i^k_k},X\ind{_a^i^k}\Big). $$
%-------------------------------------------------------------------------------------

We will derive now the reduced Tulczyjew map, which forms the basis of the Lagrangian description of gauge theories, and which is one of the main results of this paper. In subsection 3.2 we have defined $\alpha$ as a map, which is dual to $\kappa$ with respect to the evaluation (\ref{EwaluacjaJPJE}). The construction of the Tulczyjew map for field theory is therefore based on the evaluation
\begin{equation}\label{ewaluacjaTulcz}
\sJ\mathcal P\times_{\sJ C}\sJ\sV C\to\Omega^m, \qquad \langle \sj_{x}p, \sj_{x}\delta\tau  \rangle=\d \langle p, \delta\tau  \rangle(x),
\end{equation}
where $x$ is a point in $M$. Let us recall that the bundles appearing in the above evaluation have the natural identifications
$$ \sJ\mathcal P=\sJ C\times_M\sJ\bar{\mathcal P}\times_M\sJ\mathcal S, \qquad \sJ\sV C\simeq\sJ C\times_M\sJ\cV.   $$
Therefore, the evaluation (\ref{ewaluacjaTulcz}) is in fact the evaluation between elements of $\sJ\bar{\mathcal P}\times_M\sJ\mathcal S$ and $\sJ\cV$. Notice that in  $\sJ\mathcal P$ there exists a subbundle $\sJ C\times_M\sJ\barP$, which can be identified with
$$\sJ C\times\sJ\barP\simeq \sJ\mathcal P/\sJ\mathcal S.$$
Furthermore, we can project $\sJ C$ onto the second factor in (\ref{rozklJC}) and obtain the reduction of the phase bundle
$$ \sJ\mathcal P\to C\times_M\cF\times_M\sJ\barP .$$
On the other hand, in $\sJ C\times_M\mathsf J^*\cV\otimes\Omega^m$ there exists a subbundle $\mathcal K$ (\ref{podwiazkaK}), which can be subject to the symplectic reduction (\ref{podrozmaitK}). The idea of the reduction of the Tulczyjew map is represented by the diagram
$$\xymatrix{
\sJ C\times_M\sJ\barP\times_M\sJ\mathcal S \ar[r]^{ \alpha} \ar[dd]^{} & \sJ C\times_M\mathsf J^*\cV\otimes\Omega^m  \\
& \mathcal K \ar@{^{(}->}[u] \ar[d]^{}  \\
 (C\times_M\cF)\times_M\sJ\barP   \ar[r]^{\overline\alpha \qquad \qquad}    &   (C\times_M\cF)\times_M(\Omega^{m-1}\otimes\ad^*P)\times_M\barP  
}
$$
The left arrow represents the division of the phase bundle by $\sJ\mathcal S$ composed with the projection $\sJ C\to C\times_M\cF$. The right-hand side of the diagram represents the symplectic reduction of the submanifold $\mathcal K$ composed with the projection $\sJ C\to C\times_M\cF$.

Let us move to the calculations. We start with elements
$$(\sj_{x}\omega,\sj_{x}p)\in\sJ_{x}C\times\sJ_{x}\barP, \qquad  (\sj_{x}\omega, \sj_{x}\tau)\in\sJ_{x}C\times\sJ_{x}\cV.  $$
In the above identifications $\sj_{x}\omega$ is a point on the base so by means of (\ref{ewaluacjaTulcz}) we can write the evaluation between $\sj_{x}p$ and $\sj_{x}\tau$ %has a form
$$\langle \sj_{x}p, \sj_{x}\tau  \rangle=\d \langle p,  \tau  \rangle(x). $$
From now on we will abuse our notation and we will denote by $p$ a section of the bundle $\barP$ instead of $\mathcal P$. Let us introduce a basis $(e_i)$ in $\ad_{x}P$ and the dual basis $(e_*^i)$ in $\ad_{x}^*P$. We have
$$p=p\ind{_a}\otimes e_*^a, \quad p_a\in\Omega^{m-2},$$
$$\tau=\tau^a\otimes e_a, \quad \tau^a\in\Omega^1.$$
The evaluation between those elements reads 
$$\d \langle p,  \tau  \rangle(x)=\d( p\ind{_a}\wedge\tau^a)= \d p\ind{_a}\wedge\tau^a+ (-1)^{m-2}p\ind{_a}\wedge\d\tau^a=  \langle\d p,\tau \rangle+(-1)^m\langle p, \d\tau\rangle.$$
Since $(\sj_{x}\omega,\sj_xp)$ contains information about the value  $\omega(x)$, we can decompose above expression in a following way
$$\langle\d p, \tau  \rangle+(-1)^m\langle p, \d\tau  \rangle=\langle \d p,\tau \rangle+{\blue (-1)^{m}\langle p,[\omega\wedge\tau]\rangle}+(-1)^{m}\langle p,\d\tau \rangle-{\blue (-1)^{m}\langle p,[\omega\wedge\tau]\rangle}. $$
Furthermore, we have
$$   \langle p,\d\tau \rangle+\langle p,[\omega\wedge\tau]\rangle=  \langle p,\d\tau+[\omega\wedge\tau]\rangle= \langle p,D_\omega\tau \rangle,   $$
$$ \langle \d p,\tau \rangle-(-1)^{m}\langle p,[\omega\wedge\tau]\rangle=\langle \d p,\tau \rangle-\langle \ad^*(\omega)\wedge p,\tau\rangle=\langle D^\#_\omega p,\tau\rangle,  $$
where in the last equality we have used the identity
$$\langle p,[\omega\wedge\tau]\rangle= \langle p,\ad(\omega)\wedge\tau\rangle=\langle p_c\otimes e_*^c,\omega^a\wedge\tau^b\otimes\ad(e_a)(e_b)\rangle=p_c\wedge\omega^a\wedge\tau^b\otimes\langle e_*^c,\ad(e_a)(e_b)\rangle=
$$
$$= p_c\wedge\omega^a\wedge\tau^b\otimes\langle \ad^*(e_a)e_*^c,(e_b)\rangle= (-1)^{m-2}\omega^a\wedge p_c\wedge\tau^b\otimes\langle \ad^*(e_a)e_*^c,(e_b)\rangle= $$
$$=(-1)^m\langle\omega^a\wedge p_c\otimes \ad^*(e_a)e_*^c, \tau^b\otimes e_b\rangle=(-1)^m\langle\ad^*(\omega)\wedge p,\tau\rangle. $$
In the above formula $\ad(\omega)\wedge\tau$ is the particular case of the expression (\ref{rhoprim}) for $\rho$ being the adjoint representation. Notice that $\omega$ and $\tau$ can be understood as $\mg$-valued forms on $P$, so that the expression $\ad(\omega)\wedge\tau$ in this identification is the exterior product of the $\mg$-valued forms on $P$ given by \ref{nawiasgradowany}. Finally, we obtain
\begin{equation}\label{obcietaewaluacja}
\langle \sj_{x}p, \sj_{x}\tau  \rangle=\d \langle p,  \tau  \rangle(x)=    \langle D^\#_\omega p,\tau\rangle +(-1)^m\langle p,D_\omega\tau \rangle.  
\end{equation}
Notice that the above formula does not depend on the entire $\sj_{x}\tau$ but only on its projection onto $\cV\times_M\cF$. We interpret (\ref{obcietaewaluacja}) as the evaluation of the pair $( D^\#_\omega p,(-1)^mp)$ with $(\tau, D_\omega\tau)$. Combined with the fact that the map $\overline\kappa$ has form \ref{zredukkappa} we obtain that the reduced $\alpha$ reads
$$\overline\alpha:(C\times_M\cF)\times_M\sJ\barP\to (C\times_M\cF)\times(\Omega^{m-1}\otimes_M\ad^*P)\times_M\barP,$$
\begin{equation}\label{redukalpha}
\Big( \omega, F_\omega, \sj_{x}p  \Big) \longmapsto   \Big(\omega, F_\omega, D^\#_\omega p, (-1)^mp \Big).   
\end{equation}
In the above formula we have used isomorphism
$$\sV^*(C\times_M\cF)\otimes_M\Omega^m\simeq (C\times_M\cF)\times(\Omega^{m-1}\otimes_M\ad^*P)\times_M\barP.   $$

\section{Hamiltonian description}

In this section we will subject the Hamiltonian formalism to a similar reduction as the one performed in section 6.
The Hamiltonian side of gauge theories is strongly related to the dual bundle of $A=B\bar\oplus_NV_2$. Therefore, it is an important question in this context what is the picture of the vector-affine structure on the dual side. Before we can analyse Hamiltonian formalism we have to consider this issue in detail.

Since the constructions presented here are quite complicated from the technical point of view, for the clarity of presentation %we will proceed similarly as in the section \ref{TTpola}, i.e. 
we will consider first some preparatory constructions concerning general vector-affine products and then we will apply them to the bundle $\sJ C$. More precisely, in section \ref{ssec:7.1} we will consider the dual side of $A=B\bar\oplus_NV_2$ and in \ref{ssec:7.2} we will apply these results to $\sJ C$ and reduce the Hamiltonian formalism.%Hamiltonian formalism in gauge theories.

\subsection{The dual side of vector-affine product}\label{ssec:7.1}

In section (\ref{sec:1iloczyny}) we were considering an affine bundle $A\to N$ modelled on a vector bundle $V\to N$, which is the direct sum of bundles $V_1\to N$ and $V_2\to N$. Furthermore, we assumed that there exists in $A$ an affine subbundle $B\to N$ modelled on $V_1\to N$. Then, we have the decomposition \ref{decomposition1}, namely
%Then, each element $a\in A$ may be uniquely decomposed on the pair $(b,v_2)$, where $b\in B$, $v_2\in V_2$. In this way we have obtained the decomposition \ref{decomposition1}, namely
\begin{equation}\label{rozkladA}
A=B\bar\oplus_NV_2.
\end{equation}
Let us consider affine maps on $A\to N$ having the structure (\ref{rozkladA}). We denote by $U\to N$ a one-dimensional vector bundle over $N$ and by $\Aff(A, U)$ the set of all affine maps $T:A\to U$. Recall that $\Aff(A, U)$ is a vector bundle over $N$ and an affine bundle over $V^*\otimes_NU$ with the model bundle $V^*\otimes_NU\times U\to V^*\otimes_NU$ \cite{KG}. Let
$$T:A\to U $$
be an affine map and
$$\bar T:V\to U  $$
the linear part of $T$. The bundle $V$ is the direct sum of the bundles $V_1$ and $V_2$, which means that $\bar T$ is the direct sum of two linear maps
$$\bar T=\bar T_1+\bar T_2, \quad \bar T_1:=\bar T|_{V_1}, \quad \bar T_2:=\bar T|_{V_2}. $$
On the other hand, using the decomposition (\ref{rozkladA}) for $a=(b,w_2)$ we obtain
$$T(a)=T(b+w_2)=T(b)+\bar T(w_2).  $$
%oraz
%$$T(a+v)=T[(b+w_2)+v_1+v_2]=T(b)+\bar T(v_1)+\bar T(w_2)+\bar T(v_2)=T(b)+\bar T_1(v_1)+\bar T_2(w_2+v_2).  $$
It turns out that each affine map can be decomposed on two maps. The first one is an affine map
$$T_1:B\to U, \quad T_1:=T|_B,$$
with the linear part $\bar T_1$. The second one is a linear map
$$\bar T_2:V_2\to U. $$ 
From the above relations we obtain the following decomposition of the bundle $\Aff(A, U)\to N$
\begin{equation}\label{rozkladAff}
\Aff(A, U)=\Aff(B, U)\oplus_N(V_2^*\otimes_N U).
\end{equation}
The bundle $\Aff(B, U)\to V_1^*\otimes_N U$ is an affine bundle modelled on the trivial bundle $(V_1^*\otimes_N U)\times U\to V_1^*\otimes_N U$. %Jest to podwiązka afiniczna w $\theta:\Aff(A, U)\to V^*\otimes_N U$. 
The addition of the elements from $\Aff(A, U)$ and the elements of the model bundle in the decomposition (\ref{rozkladAff}) reads 
\begin{equation}\label{dodawanmodelow}
\qquad \qquad \qquad \qquad T+u=(T_1+u,\bar T_2),  \qquad  T\in\Aff(A, U), \quad  u\in (V_1^*\otimes_N U)\times U  .
\end{equation}
Our goal now will be the reduction of the map $R_A$ presented in \cite{KG} with respect to (\ref{rozkladAff}). Recall that the map $R_A$ reads %{\red (you have to attach R somewhere!)}
\begin{equation}
 R_A:\sT^*A\otimes_AU\to \mathsf P\Aff(A,U), \quad  (x^i,f^\alpha,\sigma_i,\psi_\alpha)\to (x^i,\psi_\alpha,\sigma_i,-f^\alpha)
\end{equation}
and it comes from the map $\wdt R_A$
\begin{equation}\label{symplekttildeR}
\wdt R_A:\sT^*\Aff(A,U)\otimes_{\Aff(A,U)}U\to \sT^*A\otimes_AU, \quad (x^i,\varphi_\alpha,r,\sigma_i,f^\alpha,-1) \to (x^i,-f^\alpha,\sigma_i,\varphi_\alpha),
\end{equation} 
%$$ \wdt R_A:  \sT^*(\Aff(B, U)\oplus_N(V_2^*\otimes_N U))\otimes_{\Aff(A,U)}U \to \sT^*(B\bar\oplus_NV_2)\otimes_AU. $$
which can be constructed by means of the evaluation
$$A\times_N\Aff(A,U)\to U,   $$
$$ (a,T)\longmapsto T(a).  $$
Notice that for the bundle $V_2\to N$ there exists a canonical isomorphism \cite{GU,KU}
$$R_{V_2}:\sT^*(V_2^*\otimes_N U)\otimes_{(V_2^*\otimes_N U)}U\to \sT^*V_2\otimes_{V_2}U,$$ 
which comes from the evaluation
$$ V_2\times_NV_2^*\otimes_NU\to U.  $$
%with values in $U$. 
We will show now that the reduction of $\wdt R_A$ with respect to (\ref{rozkladAff}) naturally leads to $R_{V_2}$. % how geometrically derive from, when $A=B\bar\oplus_NV_2$.
From the definition of $\wdt R_A$ we have  %Using (\ref{symplekttildeR}) we can write
\begin{equation}
 \wdt R_A: \sT^*(\Aff(B, U)\oplus_N(V_2^*\otimes_N U))\otimes_{\Aff(A,U)}U \to \sT^*(B\bar\oplus_NV_2)\otimes_AU.
\end{equation} 
Let us introduce coordinates $(x^i,b^a)$ in $B$ and $(x^i,v^\alpha)$ in $V_2$. Then, we have induced coordinates $(x^i,b^a,v^\alpha)$ in $B\bar\oplus_NV_2$.

In $\sT(B\bar\oplus_NV_2)$ we have the subbundle of vectors tangent to the fibers of the bundle $B\to N$. We will denote this subbundle by $\Delta B$ and in coordinates it reads $(x^i,b^a,v^\alpha, 0, \dot b^a, 0)$.
%Z definicji składa się ona z wektorów stycznych do krzywych $t\to(b+tX,v)\in B\bar\oplus_NV_2$, tzn. z elementów postaci $(b,v,X,0)$. 
Consider now the annihilator of $\Delta B$ in $\sT^*(B\bar\oplus_NV_2)$, which will be denoted by $\Delta B^o$. In coordinates $\Delta B^o$ has a form $(x^i,b^a,v^\alpha, \sigma_i, 0, \phi_j^V)$. The annihilator $\Delta B^o$ by definition consists of those covectors in $\sT^*(B\bar\oplus_NV_2)$, which come from functions constant along the fibers of $B$. These covectors may be identified with elements of $\sT^*V_2$, which means that there exists a natural projection 
$$\sT^*(B\bar\oplus_NV_2)\supset\Delta B^o \to\sT^*V_2, \qquad (x^i,b^a,v^\alpha, \sigma_i, 0, \phi_j^V)\longmapsto (x^i, v^\alpha, \sigma_i, \phi_j^V).  $$
A similar construction may be applied to $\sT\Aff(A,U)$. From \ref{rozkladAff} we have
$$\sT\Aff(A,U)=\sT(\Aff(B, U)\oplus_N(V_2^*\otimes_N U)).$$
Let us introduce coordinates $(x^i,\phi_j^B,r)$ in $\Aff(B,U)$, $(x^i,\phi_j^V)$ in $V^*_2\otimes_NU$ and $(x^i,\phi_j^B,\phi_k^V,r)$ in $\Aff(B,U)\bar\oplus_NV_2$. 
% (NA WYPADEK POPRAWEK) Tutaj możliwe, że trzeba będzie wziąć styczne do $\Aff(B,U)\to V^*_1\times U$ i jednocześnie leżące w K_A. To wtedy wyjdzie ten -1 we współrzędnych
Now we can proceed similarly as in the case of $\sT(B\bar\oplus_NV_2)$. In $\sT\Aff(A,U)$ we have the subbundle of vectors tangent to the fibers of the bundle $\Aff(B,U)\to N$, which we denote by $\Delta\Aff(B,U)$, and its annihilator in $\sT^*\Aff(A,U)$ denoted by $\Delta\Aff(B,U)^o$. The subbundle $\Delta\Aff(B,U)^o$ can be projected onto $\sT^*(V_2\otimes_NU)$ by means of the map
$$\sT^*(\Aff(B, U)\oplus_N(V_2^*\otimes_N U))\supset\Delta\Aff(B,U)^o\to\sT^*(V^*_2\otimes_NU),$$
$$ (x^i,\phi_j^B,\phi_k^V,r,\sigma_i, 0,v^\alpha, 0)\longmapsto (x^i,\phi_k^V,\sigma_i,v^\alpha).  $$
Using both projections we can perform the reduction of the map $R_A$, which is presented on the diagram
\begin{equation}
\xymatrix{
   \sT^*(\Aff(B, U)\oplus_N(V_2^*\otimes_N U))\otimes_{\Aff(A,U)}U  \ar[rr]^{\qquad \qquad\quad \wdt R_A} && \sT^*(B\bar\oplus_NV_2)\otimes_AU  \\
   \Delta\Aff(B,U)^o \ar@{^{(}->}[u]   \ar[d]^{}  &&  \Delta B^o \ar@{^{(}->}[u]    \ar[d]^{}   \\
  \sT^*(V_2^*\otimes_N U)\otimes_{(V_2^*\otimes_N U)}U  &&  \ar[ll]_{\qquad\quad  R_{V_2}}    \sT^*V_2\otimes_{V_2}U
}
\end{equation}
From the above diagram we obtain the reduced version of the map $R_A$
\begin{equation}\label{RV2}
R_{V_2}: \sT^*V_2\otimes_{V_2}U \to\sT^*(V_2^*\otimes_N U)\otimes_{(V_2^*\otimes_N U)}U , \quad (x^i,v^\alpha,\sigma_i,\phi_j^V)\to (x^i,\phi_j^V,\sigma_i,-v^\alpha). 
\end{equation}
%{\blue --------------------------------------
%Dalej dać opis hamiltonowski
%$$\wdt R_{V_2}:\sT^*(V_2^*\otimes_N U)\otimes_{(V_2^*\otimes_N U)}U\to \sT^*V_2\otimes_{V_2}U.$$
%Zauważmy, że $\wdt R_{V_2}$ jest szczególnym przypadkiem (\ref{IzomRF}) z tą różnicą, że za zbiór wartości ewaluacji $F$ i $F^*$ przyjęliśmy $U$ zamiat $\mathbb R$. 
%$$R_{V_2}:\sT^*(V_2^*\otimes_N U)\otimes_{(V_2^*\otimes_N U)}U\to \sT^*V_2\otimes_{V_2}U.$$
%}
Recall that according to \cite{KG} the Hamiltonian in classical field theory is represented by the family of functions
\begin{equation}\label{Hamfam}
H:A\times_N\Aff(A,U)\to U
\end{equation}
parametrised by points in $A$. As a result of the reduction described above the bundle $\Aff(A,U)$ has been replaced by the bundle $V^*_2\otimes_NU$, which is a vector bundle. It implies that the derived Hamiltonian side, in a sharp contrast to the general case, is rather linear in its nature than an affine like. In the reduced description the Hamiltonian \ref{Hamfam} reduces to the family of functions
\begin{equation}\label{reducedHam}
 H: V_2\times_NV^*_2\otimes_NU\to U,
\end{equation}
parametrised by points in $V_2$.

%{\red  Wszystko  jest ok (tylko relacja między RA i tilde RA i pytanie o lagranżjan na JC) }

\subsection{Reduced Hamiltonian formalism}\label{ssec:7.2}

Recall that according to ({\ref{Rdladżetów}}) the map $R_{\sJ C}$ reads
$$R_{\sJ C}:\sV^*\sJ C\otimes_M\Omega^m\to \mathsf P\sJd C. $$
We will use now the constructions described in the previous subsection to reduce the above map and, subsequently, to reduce the entire Hamiltonian description of gauge fields. We will be interested in the case 
\begin{eqnarray*}
A&=&\sJ_xC,\\
B&=&(\J2 P/G)_x, \\
V_1&=&C_x\times(\vee^2\sT_x^*M\otimes\ad_xP),\\
V_2&=&C_x\times\mathcal F_x,\\ 
N&=&C_x,  \\
U&=&\Omega^m_x.  \\
\end{eqnarray*}
From the above we obtain immediately
\begin{eqnarray*}
\Aff(A,U)&=&\sJd_xC,\\
\Aff(B,U)&=&(\mathsf J{^2} P/G)^\dagger_x,\\
V_1^*\otimes_N U&=&C_x\times(\vee^2\sT_xM\otimes\ad^*_xP)\otimes\Omega_x^m,   \\
V_2^*\otimes_N U&=&C_x\times\mathcal F^*_x\otimes\Omega_x^m\simeq C_x\times\barP_x,   \\
\end{eqnarray*}
where by $(\mathsf J{^2} P/G)^\dagger_x$ we have denoted the set of affine maps on $(\mathsf J{^2} P/G)_x$ with values in $\Omega^m$. 

From (\ref{RV2}) we have that the map $R_{V_2}$ reads
$$  R_{V_2} : \sT^*( C_x\times\cF_x)\otimes\Omega_x \to  \sT^*( C_x\times\barP_x)\otimes\Omega_x.  $$
Notice that we have isomorphisms $\sT^*(C_x\times\cF_x)\simeq\sV^*_x(C\times\cF)$ and $\sT^*(C_x\times\barP_x)\simeq\sV^*_x(C\times\barP)$. Taking above identifications point by point in $M$ we obtain a map
$$ R_{V_2} : \sV^*(C\times_M\cF)\otimes\Omega^m \to \sV^*( C\times_M\barP)\otimes_M\Omega^m.$$
Notice that the relations 
\begin{eqnarray*}
\sV^*C&=&C\times_M\sT M\otimes_M\ad^*P,\\
\sV^*(C\times_M\cF)&=&C\times_M\cF\times_M\sT M\otimes_M\ad^*P\times_M\cF^*,\\
\sV(C\times_M\barP)&=& C\times_M\barP\times_M\sT^*M\otimes_M\ad P\times_M\barP^*   
\end{eqnarray*}
imply the following identifications for the domain 
$$ \sV^*\Big( C\times_M\mathcal F\Big)\otimes_M\Omega^m\simeq \Big( C\times_M\mathcal F\Big)\times_M \Big( \Omega^{m-1}\otimes_M\ad^*P\Big)\times_M\barP $$
and pre-domain
$$\sV^*\Big( C\times_M\barP\Big)\otimes\Omega^m\simeq   \Big( C\times_M\barP  \Big)\times_M \Big(\sT M\otimes_M\ad^*P\otimes\Omega^m\Big)\times_M\Big(\mathfrak X^{m-2}\otimes_M\ad P\otimes\Omega^m  \Big)\simeq  $$
$$\simeq  \Big( C\times_M\barP\Big)\times_M \Big(\Omega^{m-1}\otimes_M\ad^*P\Big)\times_M\cF  $$
of the map $R_{V_2}$. In the above formula we have denoted by $\mathfrak X^{m-2}$ the space of $(m-2)$-tangent vectors to $M$.

From now on we will use the notation $\overline R:=R_{V_2}$. Let us introduce coordinates $(x^i,A\ind{^a_j})$ in $C$, $(x^i,F\ind{^a_j})$ in $\cF$, $(x^i,\sigma\ind{^a_j})$ in $\Omega^{m-1}\otimes_M\ad P$ and $(x^i,p\ind{^a_j})$ in $\barP$. Then, the map $\overline R$ reads
$$\overline R:\Big( C\times_M\mathcal F\Big)\times_M \Big( \Omega^{m-1}\otimes_M\ad^*P\Big)\times_M\barP \to  \Big( C\times_M\barP\Big)\times \Big(\Omega^{m-1}\otimes_M\ad^*P\Big)\times\cF,$$
$$ (x^i,A\ind{^a_j},F\ind{^b_k}, \sigma\ind{^a_j}, p\ind{^a_j}) \longmapsto  (x^i,A\ind{^a_j},p\ind{^a_j}, \sigma\ind{^a_j},-F\ind{^b_k}). $$
From the above formula we can recognise the geometric version of $\overline R$, which reads
%Patrząc na postać odwzorowania $R_{V_2}$ we współrzędnych i korzystając z powyższych identyfikacji rozpoznajemy natychmiast geometryczną postać odwzorowania $R_{V_2}$. From now on we will use the notation $\overline R:=R_{V_2}$. Finally we obtain
$$ \overline R: \Big( C\times_M\mathcal F\Big)\times_M \Big( \Omega^{m-1}\otimes_M\ad^*P\Big)\times_M\barP \to  \Big( C\times_M\barP\Big)\times \Big(\Omega^{m-1}\otimes_M\ad^*P\Big)\times\cF,  $$
\begin{equation}\label{redukR}
 (\omega, F, b , X )\longmapsto (\omega, X, b, -F).
\end{equation}
The reduced map $\beta$ will be denoted by $\overline\beta$ and we define it as the composition of maps (\ref{redukalpha}) and (\ref{redukR}), i.e. $\overline\beta=\overline R\circ\overline\alpha$. After some straightforward calculations we obtain
\begin{equation}\label{redbeta1}
\overline\beta:(C\times_M\cF)\times\sJ\barP\to \Big( C\times_M\barP \Big)\times \Big(\Omega^{m-1}\otimes_M\ad^*P\times\cF\Big),
\end{equation}
$$\Big(\omega, F,\sj p\Big)\longmapsto\Big(\omega, (-1)^mp, D_\omega^\#p , -F \Big).  $$
Let us move to the description of the dynamics. From (\ref{reducedHam}) we have that the reduced Hamiltonian in gauge theories is the family of $\Omega^m$-valued  functions
\begin{equation}\label{redukham}
\bar H:C\times_M(\cF\times_M\barP)\to\Omega^m, \quad \bar H(\omega, F, p)=\bar L(\omega, F)-\langle F,p \rangle  
\end{equation}
parametrised by elements in $\cF$. In some cases this family can be reduced to a single function
$$\bar H:C\times_M\barP\to\Omega^m.  $$
The dynamics on the Hamiltonian side is given by the formula
$$\mathcal D=\overline\beta^{-1}(\d\bar H(\barP)).   $$

%{\red maybe write what is the dynamics? But on the other hand its written in next section}
%{\red 
%\subsection{Wersja we współrzędnych - do wyrzucenia}
%}
%$$\overline R: (C\times_M\cF)\times_M(\Omega^{m-1}\otimes\ad^*P)\times_M\barP  \to (C\times_M\barP)\times_M(\Omega^{m-1}\otimes\ad^*P)\times_M\cF,$$
%$$(\omega, F, b , X )\longmapsto (\omega, X, b, F) .  $$
%Odwzorowanie $\overline\beta$ przyjmuje wówczas postać
%$$ \overline\beta:(C\times_M\cF)\times\sJ\barP\to \Big( C\times_M\barP \Big)\times \Big(\Omega^{m-1}\otimes_M\ad^*P\times\cF\Big).  $$
%$$\Big(\omega, F,\sj p\Big)\longmapsto\Big(\omega, (-1)^mp, D_\omega^\#p , F \Big).  $$

%Podrozdział Yang-Mills ma powtórzenia z tym co pisałem w teoriach z cechowaniem}

\section{Tulczyjew triple for gauge theories}\label{sec:5.4}

The reduced Lagrangian and Hamiltonian descriptions constitute together the reduced Tulczyjew triple, which is presented on the diagram
\begin{equation} \hspace*{-2.3cm} \qquad \qquad \qquad
\xymatrix@C-23pt{
{\red D} \ar@{^{(}->}[d] &&&&&&&&\\
(C\times_M\cF)\times_M\sJ\barP  \ar[rrrrd]^{\quad pr_{ C\times_M\cF}}  \ar[rrrrrrrr]^{\overline\alpha\qquad \qquad}&&&&&&&&
 (C\times_M\cF)\times_M(\Omega^{m-1}\otimes_M\ad^*P)\times_M\barP    \ar[lllld]_{ pr_{ C\times_M\cF}\qquad} \\
&&&&    C\times_M\cF   \ar@/_1pc/[urrrr]_{{\color{red}{\mathsf d}L}}  &&&& \\
}
\end{equation} 
for the Lagrangian side and on the diagram
\begin{equation} \hspace*{-2.3cm} \qquad \qquad \quad
\xymatrix@C-23pt{
 &&&&&&&&{\red D} \ar@{^{(}->}[d]\\
\Big( C\times_M\barP \Big)\times_M\Big(\Omega^{m-1}\otimes_M\ad^*P\times_M\cF\Big) \ar[rrrrd]_{pr_{ C\times_M\barP}}&&&&&&&&
(C\times_M\cF)\times_M\sJ\barP  \ar[llllllll]_{\qquad \qquad\overline\beta}   \ar[lllld]^{pr_{ C\times_M\barP}} \\
&&&&    C\times_M\barP    &&&& \\
}
\end{equation} 
for the Hamiltonian side.

The reduced Tulczyjew triple allows to describe the dynamics of gauge theories, with Lagrangian and Hamiltonian that does not depend on the entire first jet of a gauge field but only on the value of the connection and curvature in a given point. The dynamics is a subset in the space $(C\times_M\cF)\times_M\sJ\barP$ and its solutions are sections of the reduced phase bundle $C\times_M\barP\to C$.

Notice that the process of the dynamics generation is in our description extremely simple from the conceptual point of view. Let us stress that this description is also independent on the regularity of the system, which is a very important feature in physical applications of gauge field theories. We recall that most of the physically interesting field theories belong to the class of nonregular systems, including such important examples like electrodynamics or Yang-Mills theories, which play the fundamental role in the theory of elementary interactions. On the Lagrangian side the dynamics is given by the formula 
$$ D:= \overline\alpha^{-1}\circ\d^{\mathsf v}L(C\times_M\cF). $$
On the Hamiltonian side, when Hamiltonian reduces to a single function, the dynamics reads
$$\mathcal D=\overline\beta^{-1}(\d\bar H(\barP)).   $$
The relation between Lagrangian and Hamiltonian of the system is given by formula (\ref{redukham}). From the diagram of the reduced Tulczyjew triple we can easily derive the Legendre map for gauge theories. The vertical differential of the Lagrangian is a map 
$$ \d^{\mathsf v}L: C\times_M\cF\to  (C\times_M\cF)\times_M(\Omega^{m-1}\otimes_M\ad^*P)\times_M\barP.  $$
Composing above map with the projection on $\barP$ in the last element we obtain a map 
$$\lambda: \cF\to\barP, \qquad F\longmapsto pr_{\barP}\circ\d^{\mathsf v}L(F),  $$
which we call the {\it Legendre map} for gauge field theories.

\section{Example: Yang-Mills theory}\label{sec:5.5}

%Przedstawimy teraz zastosowanie powyższego formalizmu do opisu klasycznej wersji teorii Yanga-Millsa, stanowiącej podstawę współczesnego opisu odziaływań elementarnych. Zaczniemy od wprowadzenia pewnych pojęć związanych z metryką na rozmaitości i wiązce dołączonej, które potrzebne będą do skonstruowania lagranżjanu Yanga-Millsa. 

%\subsection{Przykład: teorie Yanga-Millsa}

We will present now the example of application of the reduced Tulczyjew triple, which is the dynamics of Yang-Mills theory \cite{YangMills,IS}. Let us notice that in the physical literature this class of theories is usually considered on the flat Minkowski space. Our approach is more general, whereas it allows to formulate Yang-Mills theory on any smooth manifold equipped with a metric $g$. Lagrangian of a free Yang-Mills field reads
$$ \bar L:C\times_M\cF\to\Omega^m,  \quad \bar L(\omega,F)=\frac{1}{2}K_{ab}F^a\wedge\star F^b,  $$
where $F=F^a\otimes e_a=F\ind{^a_i_j}\d q^i\wedge\d q^j\otimes e_a$, $K$ is a scalar product on $\ad P$ and $\star$ is the Hodge star on $M$ coming from the metric $g$. We denote by $\Xi$ a projection on the second factor in (\ref{rozklJC}), i.e. 
$$\Xi:\sJ C\to C\times_M\cF.$$
By means of $\Xi$ we can construct the Lagrangian 
$$L:\sJ C\to\Omega^m, \quad  L=\bar L\circ\Xi.   $$
%Z powyższego widać, że teoria Yanga-Millsa powinna dać się opisać za pomocą zredukowanej trójki Tulczyjewa dla teorii z cechowaniem. Spróbujmy, więc znaleźć dynamikę dla $L$. 
We start with calculating the vertical differentials
$$\d^{\mathsf v}L:\sJ C\to \sV^*\sJ C\otimes\Omega^m,  $$
$$\d^{\mathsf v}\bar L:C\times_M\cF\to \sV^*(C\times_M\cF)\otimes\Omega^m.  $$
Above bundles may be written in the form
$$\sV^*(C\times_M\cF)=(C\times_M\cF)\times_M \cV^*\times_M\cF, $$
$$\sV^*(C\times_M\cF)\otimes\Omega^m=(C\times_M\cF)\times_M(\Omega^{m-1}\otimes_M\ad^*P)\times_M(\Omega^{m-2}\otimes_M\ad^*P). $$
Let
$$\gamma:\mathbb R\to \sJ_mC, \qquad  t\longmapsto \sj_m\omega+t\sj_m\tau, \qquad \textrm{where} \qquad  \sj_m\omega\in\sJ_mC, \quad \sj_m\tau\in\sJ_m\cV   $$
be a curve representing a vertical tangent vector $(\sj_m\omega,\sj_m\tau)$ from $\sV_m\sJ C\simeq\sJ_mC\times\sJ_m\cV $. Acting by $\d^{\mathsf v}L$ on that vector we obtain 
$$\d^{\mathsf v}L(\sj_m\omega)(\sj_m\tau)=\ddt L(\sj_m\omega+t\sj_m\tau)=\ddt \bar L(\omega+t\tau,F_{\omega+t\tau})=  $$
$$=\ddt \Big(\frac{1}{2}K_{ab}F_{\omega+t\tau}^a\wedge\star F_{\omega+t\tau}^b\Big)=  \frac{1}{2}K_{ab}\Big(\ddt F_{\omega+t\tau}^a\Big)\wedge\star F_{\omega}^b+\frac{1}{2}K_{ab}F_{\omega}^a\wedge\star\Big(\ddt F_{\omega+t\tau}^b\Big) $$
$$= K_{ab}\Big(\ddt F_{\omega+t\tau}^a\Big)\wedge\star F_{\omega}^b. $$
The curve $F_{\omega+t\tau}$ reads
$$F_{\omega+t\tau}=\d(\omega+t\tau)+\frac{1}{2}[(\omega+t\tau)\wedge(\omega+t\tau)]=F_\omega+td\tau+t[\omega\wedge\tau]+ t^2\frac{1}{2}[\tau\wedge\tau]= $$
$$=F_\omega+tD_\omega\tau+ t^2\frac{1}{2}[\tau\wedge\tau]  $$
and by taking its derivative in $t=0$ we obtain
$$\ddt F_{\omega+t\tau}= D_\omega\tau,   $$
where $D_\omega\tau= d\tau+[\omega\wedge\tau]$ is the covariant derivative of the one-form $\tau$. Finally, we obtain
$$\d^{\mathsf v}L(\sj_m\omega)(\sj_m\tau)=K_{ab}(D\tau)^a\wedge\star F_{\omega}^b.$$ %= {\red  (D\tau | F_{\omega})\in\Omega^m. } $$
Notice that
$$ K_{ab}(D\tau)^a\wedge\star F_{\omega}^b=\langle D\tau | \tilde K(\star F_{\omega})    \rangle,  $$
which implies
\begin{equation}\label{rozniczkYM}
\d^{\mathsf v}\bar L(\omega, F)=\Big(\omega, F, 0, \star \tilde K(F)\Big).
\end{equation}
Now we can apply the reduced Tulczyjew map. For simplicity let us assume that $M$ represents a four-dimensional spacetime so that $\dim M=4$.
Then, the coefficient $(-1)^m$ in (\ref{redukalpha}) disappears and we obtain 
$$\overline\alpha: \Big( \omega(m), F(m), \sj_mp  \Big) \longmapsto   \Big(\omega(m), F(m), D^\#_\omega p(m), p(m)  \Big).   $$
From (\ref{rozniczkYM}) we obtain that the equations describing the dynamics $\mathcal D=\overline\alpha^{-1}\d^{\mathsf v}\bar L(C\times_M\cF)$ read
\begin{eqnarray}
p&=&\star\tilde K(F),\\
D_\omega^\#p&=&0.
\end{eqnarray}
Let us notice that the above relations imply an equation
$$D_\omega^\#\star\tilde K(F)=0,$$
which is a generalised Yang-Mills equation. For $M=\mathbb R^4$ equipped with the Minkowski metric we obtain the traditional Yang-Mills equation
$$D_\omega\star F=0.$$

%{\red Here write how above equation can be simplified to a traditional "flat" Yang-Mills. Additionally explain that scalar product in red above}

\section{Acknowledgments}

We gratefully acknowledge prof. Katarzyna Grabowska for the fruitful conversations and valuable comments, which definitely make this paper more correct and easy to read. 

%We gratefully acknowledge anonymous referees for their valuable comments which definitely make this paper more correct and easy to read. 

\end{document}